# Reinforcement Learning Based Approaches to Adaptive Context Caching in Distributed Context Management Systems

Reinforcement Learning Based Adaptive Context Caching

Shakthi Weerasinghe

School of Information Technology, Deakin University, syweerasinghe@deakin.edu.au

Arkady Zaslavsky

School of Information Technology, Deakin University, arkady.zaslavsky@deakin.edu.au

Seng W. Loke

School of Information Technology, Deakin University, seng.loke@deakin.edu.au

Alexey Medvedev

School of Information Technology, Deakin University, alexey.medvedev@deakin.edu.au

Amin Abken

School of Information Technology, Deakin University, a.abkenar@deakin.edu.au

Alireza Hassani

School of Information Technology, Deakin University, ali.hassani@deakin.edu.au

Performance metrics-driven context caching has a profound impact on throughput and response time in distributed context management systems for real-time context queries. This paper proposes a reinforcement learning based approach to adaptively cache context with the objective of minimizing the cost incurred by context management systems in responding to context queries. Our novel algorithms enable context queries and sub-queries to reuse and repurpose cached context in an efficient manner. This approach is distinctive to traditional data caching approaches by three main features. First, we make selective context cache admissions using no prior knowledge of the context, or the context query load. Secondly, we develop and incorporate innovative heuristic models to calculate expected performance of caching an item when making the decisions. Thirdly, our strategy defines a time-aware continuous cache action space. We present two reinforcement learning agents –a value function estimating actor-critic agent and a policy search agent using deep deterministic policy gradient method. The paper also proposes adaptive policies such as eviction and cache memory scaling to complement our objective. Our method is evaluated using a synthetically generated load of context sub-queries and a synthetic data set inspired from real world data and query samples. We further investigate optimal adaptive caching configurations under different settings. This paper presents, compares, and discusses our findings that the proposed selective caching methods reach short- and long-term cost- and performance-efficiency. The paper demonstrates that the proposed methods outperform other modes of context management such as redirector mode, and database mode, and cache all policy by up to 60% in cost efficiency.

## 1 INTRODUCTION

Internet-of-Things (IoT) has gained widespread use in the recent decades [43]. From small-scale health monitoring devices to complex industrial IoT, knowledge derived from logically inferencing big IoT data can provide valuable insights into entities and the environment they interact with [29]. We refer to this knowledge as context [1] in general and the interactions of entities to cause inter-relatedness of context [17]. IoT applications benefit immensely from providing context-aware services [1] such as smart health, and smart cities. For instance, consider a smartwatch that provide health recommendations to a user based on measured vitals. Assume the same user is now approaching a smart vehicle. The vehicle may still need to connect to the smartwatch, to appropriately adjust the cabin climate for the user. Interoperability of IoT devices, and applications is therefore important for context-awareness but restrictions in data sharing among IoT applications contradicts with this notion. Context management systems solves this problem by mediating between IoT devices and application enabling seamless access IoT context without requiring manual integration with IoT devices or applications, e.g., the Context-as-a-Service (CoaaS) platform [15]. The challenge though is to efficiently manage big IoT data and produce context in a time-critical manner using limited computing resources. Caching is widely used strategy in computer systems to minimize resource usage by reusing and repurposing cached data. But context data is unique compared to regular data such as flat files, making caching context different from traditional data caching.

According to the definition given by Dey, context is "any information that can be used to characterize the situation of an entity. An entity is a person, place, or object that is considered relevant to the interaction between a user and an application, including the user and applications themselves" [1]. There are four features of context based on this definition. Firstly, context can exceed the volume of the data since it explains a certain piece of data. Context is bigger than big data. Secondly, context is transient, e.g., the location of a moving vehicle. Thirdly, context is both time and quality critical. For instance, consider notifying a driver of a potential hazard (context) ahead. Fourthly, big IoT data used to infer context are heterogeneous, e.g., in size and format. It is non-trivial to develop a context management system that solve all the four challenges considering related work [8,22,44,45], that suffer either from scalability, complexity and/or cost.

We can discuss four differences of context caching compared to data caching relating to the scope of this paper. First, the primary objective of data caching is location transparency to minimize retrieval latency. The cached item is a copy of the original data item that is placed on a faster to access node, e.g., a dedicated web application or service. On the contrary, consider higher-level context (or in other words, derived context) based on, i.e., Context Space Theory [7] such as the situation description of a room. The inputs may be noise level, light intensity, and a video stream. The context, as output, is JSON document. Cached context is not always a fast-access copy since it can have many alternate representations [17]. Different context consumers may be interested in different aspects of the same context based on their requirements as well. Secondly, context is logically hierarchical by structure as we indicate in Figure 2 compared to a single data object, e.g., a flat file. Low-level context, e.g., context attributes, are shared between higher-level context as a result. Thirdly, caching all IoT data may not be useful or cost-efficient in context caching. For instance, compare caching the geolocation from a GPS sensor of a moving car against caching the estimated trajectory of the car as context. The latter is more space and cost-efficient considering the number of refreshes need to retain a reliable value for geo-location in cache. Fourthly, context queries are heterogenous and IoT data during ingestion lack any clear pattern [25]. Adapting these variations is necessary to ensure the cache performs efficiently. Based on the fact that IoT data are the basis for deriving context, IoT devices are referred to by context providers in the rest of the paper.

Let us consider a real-world scenario surrounding a city circle with multiple intersections as illustrated in Figure1. Assuming all the requests to the context management system are pull-based queries [14], Table 1 lists several context queries that could originate from context consumers using the Context Definition and Query Language (CDQL) [13,14].

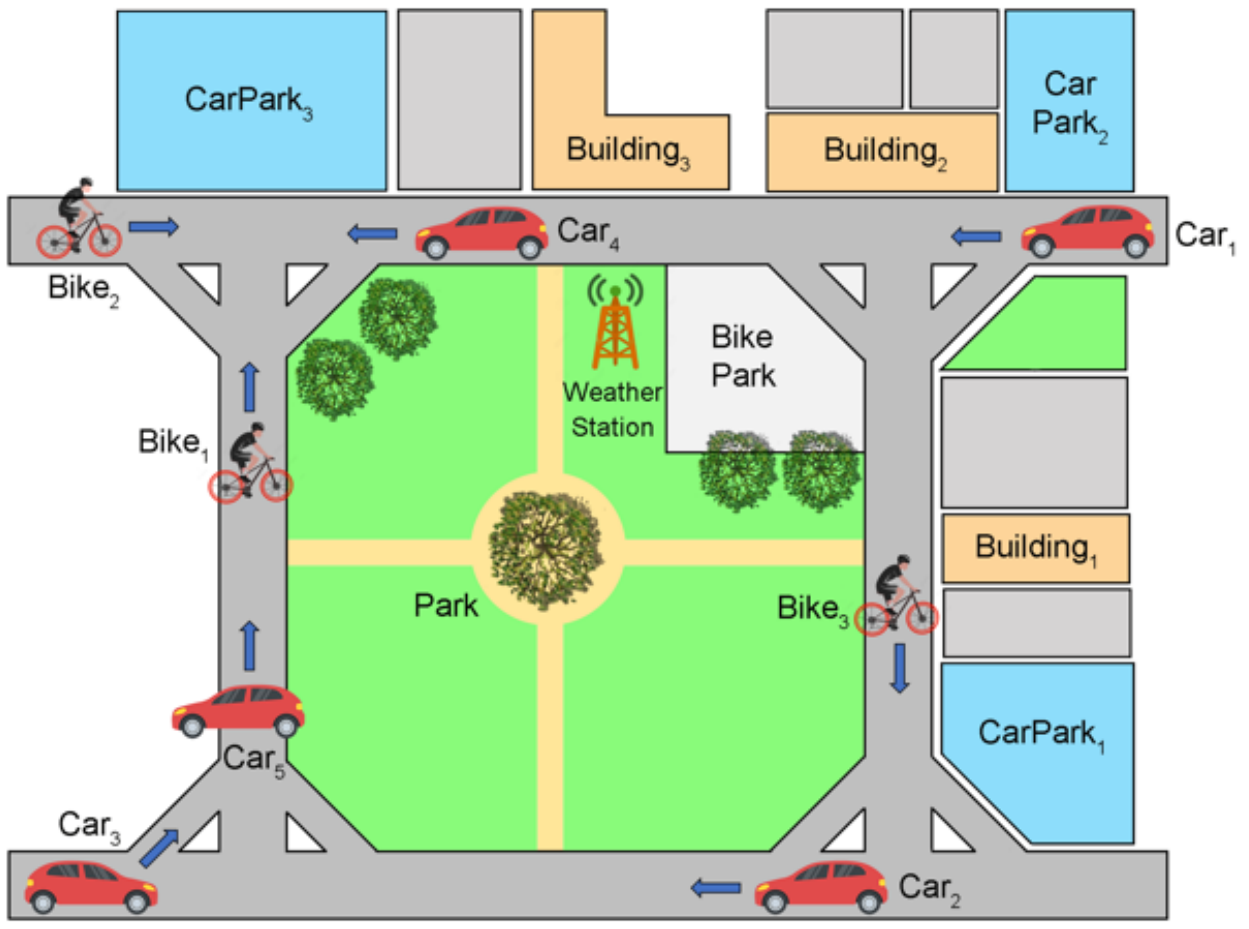

Figure 1: An intersection with multiple context providers and context consumers.

Table 1: Context queries in the motivating scenario

| Notation | Query Description |
|---|---|
| $Q_1$ | Search car parks with available slots in the area. |
| $Q_2$ | Search car parks with available slots in the area that charge less than a given amount per hour. |
| $Q_3$ | Search car parks with available slots that are less than a certain distance to the target location. |
| $Q_4$ | Search car parks with available slots given it is good for walking to the target location. |
| $Q_5$ | Check whether it is good for jogging at the park. |
| $Q_6$ | Search bike parking spots in the near vicinity of the rider. |
| $Q_7$ | Is the rider vulnerable to crash into an object at the next closest intersection? |
| $Q_8$ | Which way to turn at the next junction to avoid traffic to reach the target location? |

Context queries can be broken down into multiple sub-queries by entities at execution. For instance, the query coordinator in the Context Query Engine (CQE) coordinates the execution of the subqueries to produce the context query output in Context-as-a-Service (CoaaS) platform [15]. Figure 2 illustrates the query breakdown for $Q_4$ and $Q_5$. Blue colour is used to indicate the context query, orange colour for sub-queries, green for entities, and yellow for context attributes. The arrows indicate the logical relationship between the context items. Context in the immediate child nodes are either those required to, (i) derive the context depicted in the parent node (e.g., "*get car parks*" is required to generate context response for $Q_4$), or (ii) describe the context in the parent node (e.g., address and location attributes describe the building entity).

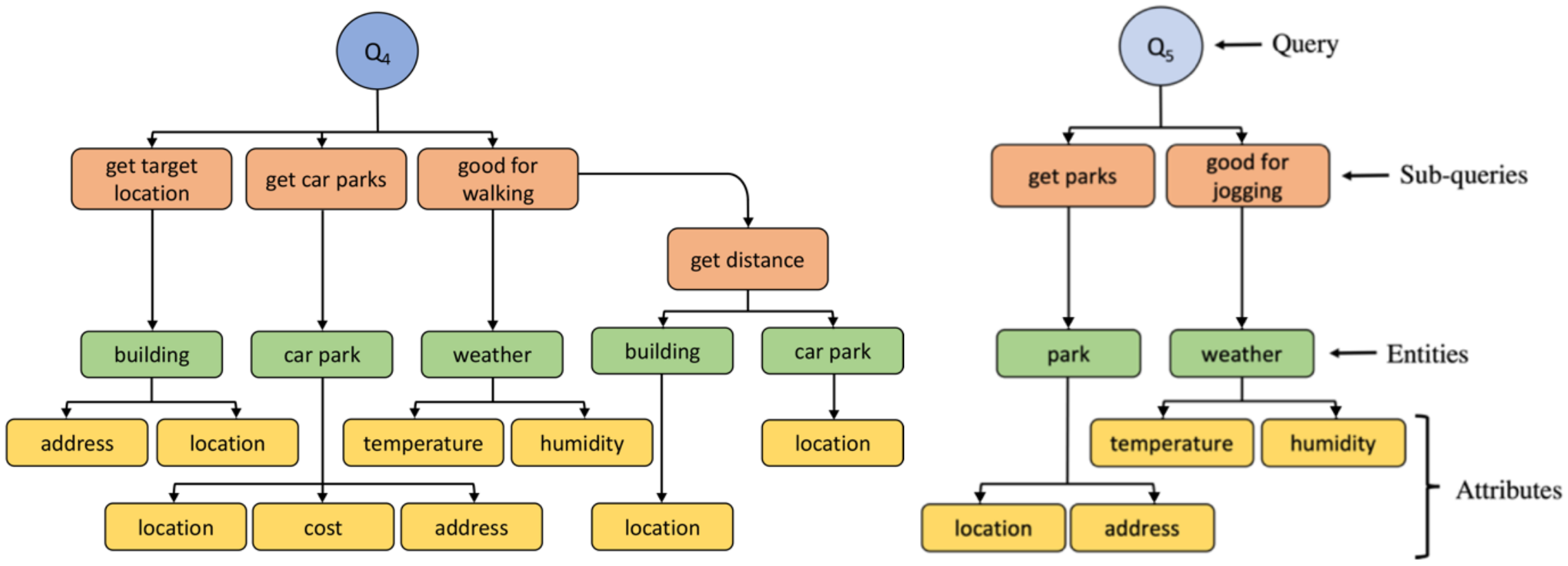

(a) (b)

Figure 2: Query breakdown for (a) $Q_4$, and (b) $Q_5$.

Notice the sub-query "*good for jogging*" share the same entities and context attributes as "*good for walking*", i.e., temperature, probability of rain attributes of weather entity. It is prudent to cache these context items for reuse and avoid redundant retrievals. The other function of context data caching is repurposing. Assume the $car_5$ executes $Q_2$ travelling close to $ar_3$ that executes $Q_1$. Retrieved data of nearby car parks for $Q_1$ could be repurposed for $Q_2$ using only a filter. But the need for each context item varies with time which is the primary motive in investigating an adaptive solution for context caching.

To the best of the authors' knowledge, this paper presents a novel approach for adaptive context caching different from traditional data caching. We also present experimental results from our prototype using a simulated context sub-query load related to the motivating scenario. The main contributions of this paper are as follows:

- We propose reinforcement learning based agents that make selective caching decisions to optimize cost and performance efficiency of adaptive context caching,
- We propose and evaluate policies, and heuristics for adaptive context caching algorithm such as eviction, and dynamic cache memory scaling,
- We evaluate the cost of responding to context queries and the Quality of Service (QoS) under multiple settings to identify the most efficient adaptive caching configuration,
- We show that our proposed algorithm performs without prior knowledge and makes the adaptive caching decisions by observing the features of the dynamic query load in real-time.

The rest of the paper is structured as follows. Section 2 discusses the background and related work. Then in Section 3, we describe our algorithms for adaptive context caching (ACOCA) using the selective caching agent, and in Section 4, the adaptive context caching algorithm is described including heuristics, and cache memory organization. Next, in Section 5, we describe the implementation of the ACOCA-A prototype, experimentations, and an extensive discussion on the observed results. Section 6 concludes and highlights the direction for future work in this area.

## 2 RELATED WORK IN ADAPTIVE CONTEXT CACHING

Caching is a popular strategy to optimize for demanding real time loads. Efficient caching has a significant impact on performance compared to accessing data via an expensive backend [10]. Zhou et al. [11] indicate a 35% reduction in perceived latency for a 1% increase in cache hit rate. One of the critical factors against reaching these efficiencies is the time-variedness of data access requests. Popularity [41], and request patterns [36] for data and queries varies temporally. Context originating from IoT devices are also transient [38,42]. Previous work investigate adaptiveness [4] based on data lifetime [38], properties of network queuing [40], popularity [27,30,31,42], and/or cost of caching [30,32,42] as features.

But there are several key drawbacks. Firstly, these are adaptive data caching solutions. We distinguished data caching from context caching above. The underlying assumptions in data caching are inapplicable to context caching. For instance, the authors in [30] define a file or data library, e.g., each sensor produces a distinct data item [27]. A similar context library cannot be defined owing to the dynamic emergence of novel context. Secondly, the optimization goals are achieved as a function of evictions. Consider, a limited sized cache using the Least Frequently Used (LFU) eviction policy. Under eviction-based optimization, an item ($i_1$) that is requested only once could evict an item ($i_0$) that was cached immediately before due to the lack of access frequency at the time $i_1$ is retrieved. But $i_1$can be a frequently accessed item resulting in at least one more eviction. So, eviction-based optimization can be both cost [12] and space inefficient due to redundant operations. Blanco et al. [6], Sadeghi et al. [30], and Chen et al. [9] selectively caching. It could avoid redundancies and help maintain consistent QoS which is unachievable with frequent evictions.

A Markov Decision Process (MDP) aimed at minimizing the expected average cost is investigated in [32]. The estimated cost approach based on remaining life neglects items with short lifetime from caching despite higher access frequency. Kiani et al. [19] provide evidence of performance advantage of dynamically resizing cache memory over a fixed cache size but need further investigation into improving the marginal results. Zhu et al. [42], Sheng et al. [31] and Nasehzadeh et al. [27] approaches MDP using Reinforcement Learning (RL) suggesting a significant reduction in response latency. But they are optimized for Quality-of-Experience (QoE) of individual consumers. Individual consumer preferences are relatively stable allowing machine learning models to converge over several iterations compared to a variable context query stream of many consumers [28]. We can also assume the heterogeneity of context queries to be significantly greater compared to a set of user interactions with a limited set of data, i.e., data library. Defining a state space similarly to previous work in data caching for context caching is non-trivial due to this reason.

In summary, this paper identifies and address a significant research gap, which is the lack of a comprehensive solution for adaptive context caching. The primary objective of this work is to develop and test a cost-efficient adaptive context caching strategy that maximizes the QoS of context delivery to the consumer while maintaining consistency in offered QoS to the consumers. Our problem is a contradicting bi-objective optimization problem involving minimizing cost while maximizing QoS (or in other works, performance of the context management system for simplicity). In the next section, we describe our approaches by developing our selective caching algorithms and applicable heuristics.

## 3 SELECTIVE CONTEXT CACHING APPROACHES

In this section, we describe the algorithms and design specifics of the adaptive context caching (which we call ACOCA) agent in detail. The selective decision is concerned with answering two principal questions: (1) *what*, and (2) *when* to cache a context item. The main objective of selectively caching context is to maximize the performance, scalability, and profitability of context management systems by efficiently utilizing the available resources – computing, cache, memory, network, and storage. Service Level Agreements (SLA) underpin the parameters to establish the desirable QoS levels which define the constraints in this optimization problem.

The popularity of a context item could be observed under two states – cached and not cached. In this paper, Hit Rate (HR) is affected in both these states. For instance, the HR of a context attribute (CA) is the conditional probability of retrieving the values from cache conforming to the applicable freshness requirement. A context item *i* that is either unavailable or obsolete in the cache is considered a miss. Access Rate (AR) is the number of references to a context item from context queries irrespective of the cached status $(n_i)$ against the total number of context queries $(N)$ received. They can be indicated as follows where *i* is an index of a context item in $i \in \mathcal{I} := \{0,1,2,\ldots,I\}$, $HR(t)_i$ and $AR(t)_i$ are the HR and AR of context item *i* at time *t*:

$$HR(t)_i = P(Fresh|Cached)_i \qquad (1)$$

$$AR(t)_i = \frac{n_i}{N} \qquad (2)$$

The number of accesses to a cached and fresh item is $m_i = k \times n_i$ where $0 \leq k \leq 1$. A sliding time window of size W corresponding to the last W time units, e.g., W seconds, is used in this paper. *T* is the index referring to the current window. To summarize the observed and expected values of sequential data, i.e., HR, AR, in a space-efficient manner, we define $< short, mid, long >$. It refers to the relative index from the current window at which the corresponding statistic is considered. For example, consider $< 1,2,5 >$ and a queue of observed HR for the past five windows as [0,0.67,0.93,0.86,0.73]. This is summarized as [0.73,0.67,0].

An important definition in this work is the Probability of Delay (PD). It is the overall probability that a context query exceeds the maximum accepted response time $(RT_{max})$ given in a SLA, resulting in a penalty $(Pen_{del})$. Caching all context items does not guarantee $RT < RT_{max}$ due to, (i) infeasible SLAs [38], (ii) ephemeral expiry periods $(E_n)$ due to short residual lifetimes (ResiL) (refer Section 3.3), (iii) processing and/or storage overheads, etc. Therefore, identifying and deciding not to cache certain items under these circumstances, e.g., $ProcessingCost \gg RetrievalCost$, can be considered an important feature of a well-designed selective context caching algorithm.

The cached lifetime of a context item $(CL_i)$ is the period of time measured in *t* time units, i.e., seconds during which *i* resides in the cache. Delay time $(DT_i)$ is the number of windows to skip until the same *i* will be evaluated for caching. Calculations of both these values are done by the selective agent algorithm which will be further described in Section 3.1, 3.2 and 3.3.

The selective caching agents investigated in this paper can be summarized as follows in Figure 3. We presented our work using the Statistical based agent (StatAgn) in [37]. We use it to compare the RL agents in the paper. The rest of the symbols used in this paper are summarized in Table 2.

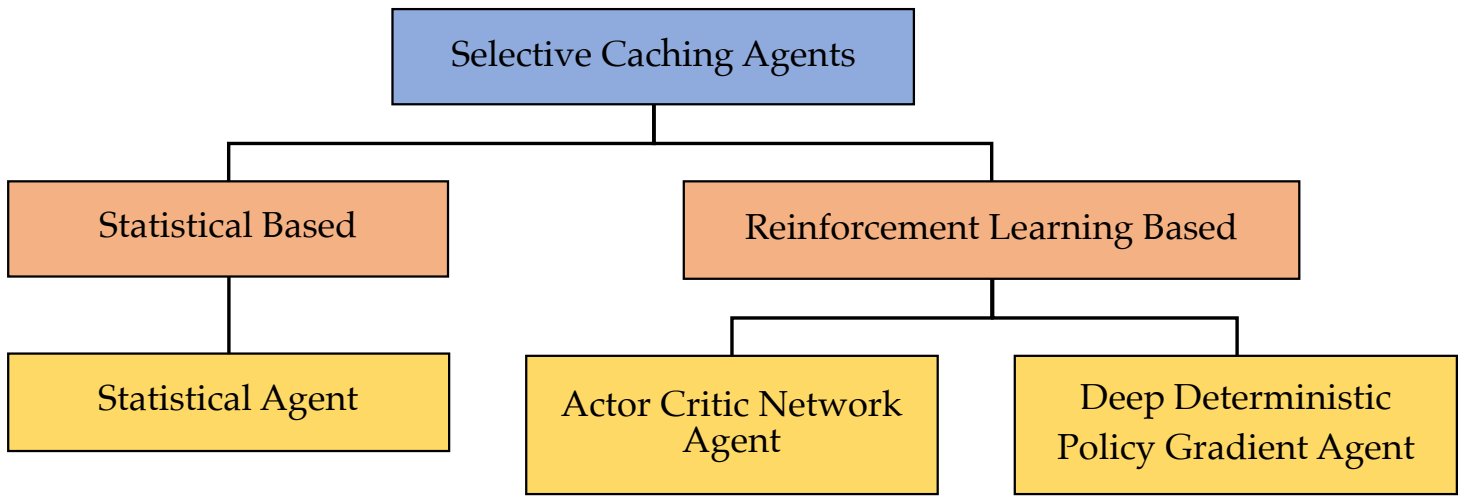

Figure 3: Relationships among entities in the motivating scenario.

Table 2: Symbols used in this paper

| Symbol | Description |
|---|---|
| $\lambda$ | Poisson request arrival rate, e.g.., 1.5/second |
| $\varepsilon$ | Exploration factor ($0 \leq \varepsilon \leq 1$) |
| $\zeta$ | Exploration factor correction |
| $\eta$ | Eviction threshold |
| $\gamma$ | Discount rate of the RL agent, i.e., 0.9 ($0 \leq \gamma \leq 1$) |
| $\alpha$ | Learning rate of the actor network, i.e., 0.0001 |
| $\beta$ | Learning rate of the critic network, i.e., 0.001 |
| $\tau$ | Factor for soft target network parameter updates |
| $SR_q$ | Sampling rate of context provider q in Q, e.g., 0.33Hz |
| $P_{DC,i}$ | $P(Delay\|Cached)_i$ – The conditional probability of getting a delayed response though the item is cached |
| $f_{thresh}$ | Minimum accepted freshness given an SLA |
| $f_{arrive}$ | Freshness of an item once retrieved |
| $Price_{res}$ | Price earned per valid context query response |
| $Cost_{re}$ | Cost of retrieving context from context providers |
| $ReL$ | Retrieve latency of a context provider |
| $L_i$ | Lifetime of a context item |
| $Cycle_{learn}$ | Learning Cycle, e.g., 20 selective caching decisions |

### 3.1 Actor Critic Agent

The first strategy adopts a value function estimation approach to RL using an actor-critic network. The iterative learning scheme allows the agent to converge to an optimal solution called the goal state. It is the distinctive feature of both the agents presented in this paper against the *StatAgn* since the latter is oblivious to long-term optimality. In fact, *StatAgn* make decisions based only on currently available statistics for the item.

We define the selective caching problem as a Markov Decision Process (MDP). A state $(s)$ as a vector containing fifteen features. They are the observed $AR(t)_i$ and $HR(t)_i$, $\mathbb{E}[AR(t)_i]$, $\mathbb{E}[HR(t)_i]$, $\overline{CL_\iota}$, average latency to retrieve, and average retrieval cost. They are indicated in Table 3. These features were chosen to reflect the (i) popularity, (ii) contributing to efficiency as a result of caching, (iii) ease of retrieval, e.g., it is easy to retrieve when the context can be retrieved at low latency and the context provider is available, and (iv) possibility to maximize cache lifetime. Observed and expected timelines are discretized using $< short, mind, long >$ tuple. The state space $(\mathcal{S})$ is continuous as a result. We have not considered size of context items as a feature in this work despite related work [20,40], assuming context are small and uniform in size. The goal is to maximize caching context items that yield a positive return to the context management system.

Table 3: Symbols used in this paper

| Notation | Description |
|---|---|
| $AR(long)_i$ | Observed AR at $T - long$ for *i.* |
| $AR(mid)_i$ | Observed AR at $T - mid$ for *i.* |
| $AR(short)_i$ | Observed AR at $T - short$ for *i.* |
| $\mathbb{E}[AR(long)_i]$ | Expected AR at $T + long$ for *i.* |
| $\mathbb{E}[AR(mid)_i]$ | Expected AR at $T + mid$ for *i.* |

| Notation | Description |
|---|---|
| $\mathbb{E}[AR(short)_i]$ | Expected AR at $T + short$ for *i*. |
| $HR(long)_i$ | Observed HR at $T - long$ for *i*. |
| $HR(mid)_i$ | Observed HR at $T - mid$ for *i*. |
| $HR(short)_i$ | Observed HR at $T - short$ for *i*. |
| $\mathbb{E}[HR(long)_i]$ | Expected HR at $T + long$ for *i*. |
| $\mathbb{E}[HR(mid)_i]$ | Expected HR at $T + mid$ for *i*. |
| $\mathbb{E}[HR(short)_i]$ | Expected HR at $T + short$ for *i*. |
| $\overline{CL_i}$ | Average cached lifetime of *i*. |
| $\overline{RL_i}$ | Average context retrieval latency of *i*. |
| $\overline{Cost_{Ret,i}}$ | Average cost of context retrieval of *i*. |

An action $(a_i)$ is defined in the action set $= \{0,1\}$ corresponding to not-caching and caching decisions for each context item *i*. The action space is $\mathcal{A} = \{a_1, a_2, \dots a_i\}$ for all $i$. Given the decision is binarily modeled, then the output is the probability to cache $i$. The caching policy as $\pi(a|s)$ accordingly. The continuous MDP model suggests that rewards realized later in time are affected by the actions taken "now". Consider that $Reward_\tau$ is the cumulative reward after $\tau$ number of windows where $\gamma$ is the discount factor. Then $Reward_\tau = \sum_{t=0}^{T}(\gamma^t \times r_{t+\tau})$ and the optimal policy $\pi^* = argmax_\pi(\mathbb{E}[Reward_\tau|\pi])$. The value function $(V^\pi)$ and Q-functions $(Q^\pi)$ for a policy can be defined as follows given $p(s'|s,a)$ is the transition probability to the new state $s'$ from *s* taking the action *a*:

$$V^\pi(s) = \sum_a \pi(a|s) \sum_{s'} p(s'|s,a)[r + (\gamma \times V^\pi(s'))] \quad (9)$$

$$Q^\pi(s,a) = \sum_{s'} p(s'|s,a)[r + \gamma \sum_{a'}(\pi(a'|s') \times Q^\pi(s',a'))] \quad (10)$$

Figure7 depicts the system design of the sub-components of the *ACAgn*. We propose to adopt the publisher-subscriber and observer patterns to handle the stream of context items for evaluating and triggering reward calculations respectively. The state translator converts a context item to the respective state which is published to the queue. Then the actor-critic network reads from the queue and makes the selective caching decision which is logged in the Decision History (DH) and Delay Time Registry (DTR) or Cache Life Registry (CLR) if the decision was not to cache and cache respectively. Context items are written to the cache memory when $a_i = 1$. Each cached context is profiled using the Cache Profiler (CP). The Event Listener listens to two types of events originating from the DHR, CLR, and/or CP. Firstly, for "calculate reward" events. The event is triggered by: (i) an item is prematurely evicted, i.e., prior to $\mathbb{E}[CL_i]$ elapsing, (ii) $\mathbb{E}[CL_i]$ has elapsed when $a_i = 1$, (iii) $\mathbb{E}[DT_i]$ has elapsed when $a_i = 0$, or (iv) the dedicated limited sized list collecting the sequence of $Ret_i$ in CP is full. Secondly, for "extend cache life" events are triggered when an item avoids eviction and has also elapsed $\mathbb{E}[CL_i]$. Accordingly, the first event causes the actor-critic network to learn and the latter to make a selective caching decision.

The output of the actor-critic network $(0 \leq v_i \leq 1)$ can be described as a measure of *fitness* to cache over the period of $\mathbb{E}[CL_i]$ defined in the feature vector since it is the probability to cache the item *i*. For instance, $v_i = 1$ suggests the item could be cached for the entire expected period and not suitable to cache when $v_i = 0$. So, $v_i = 0.5$ is an ambiguous state. Using this principle, we can map the output value to the binary cache decisions $(a_i)$ as:

$$a_i = \begin{cases} 1; if\ v_i > 0.5 \\ 0; otherwise \end{cases} \quad (3)$$

Accordingly, we define $CL_i = (v_i - 0.5) \times \overline{CL_i}/0.5$ and $DT_i = (0.5 - v_i) \times mid/0.5$.

As indicated in Figure 4, we use a delayed reward calculation scheme using event triggers and listeners to capture the actual return from the decision as opposed to calculating expected values for the reward at the decision epoch, e.g., in [16,27]. The window during which the reward is calculated is referred to as the *learning epoch*. Consider $R_{i,t+\tau}$ is the number of requests that access the context item *i* between the selective decision epoch *t* and the learning epoch after $\tau$ windows. Accordingly, $\tau = DT_i$ when $a_i = 0$ and $\tau = \min(CL_i, long, \tau_{full})$ given $a_i = 1$. $\tau_{full}$ is the time at which the dedicated limited-sized list collecting the sequence of $Ret_i$ in CP gets full. Then the reward is calculated as follows:

$$Reward = \begin{cases} f;\ a_i = 1 \\ g;\ otherwise \end{cases} \quad (4)$$

$$f = \begin{cases} \frac{\sum_{j=0}^{R_{i,t+\tau}} Ret_{i,j}}{R_{i,t+\tau}};\ if\ R_{i,t+\tau} > 0 \\ -10;\ otherwise \end{cases} \quad (5)$$

$$g = \begin{cases} \frac{\sum_{j=0}^{R_{i,t+\tau}} (Ret(t)_i - \mathbb{E}[Ret(t)_i])}{R_{i,t+\tau}};\ if\ R_{i,t+\tau} > 0 \\ 5;\ otherwise \end{cases} \quad (6)$$

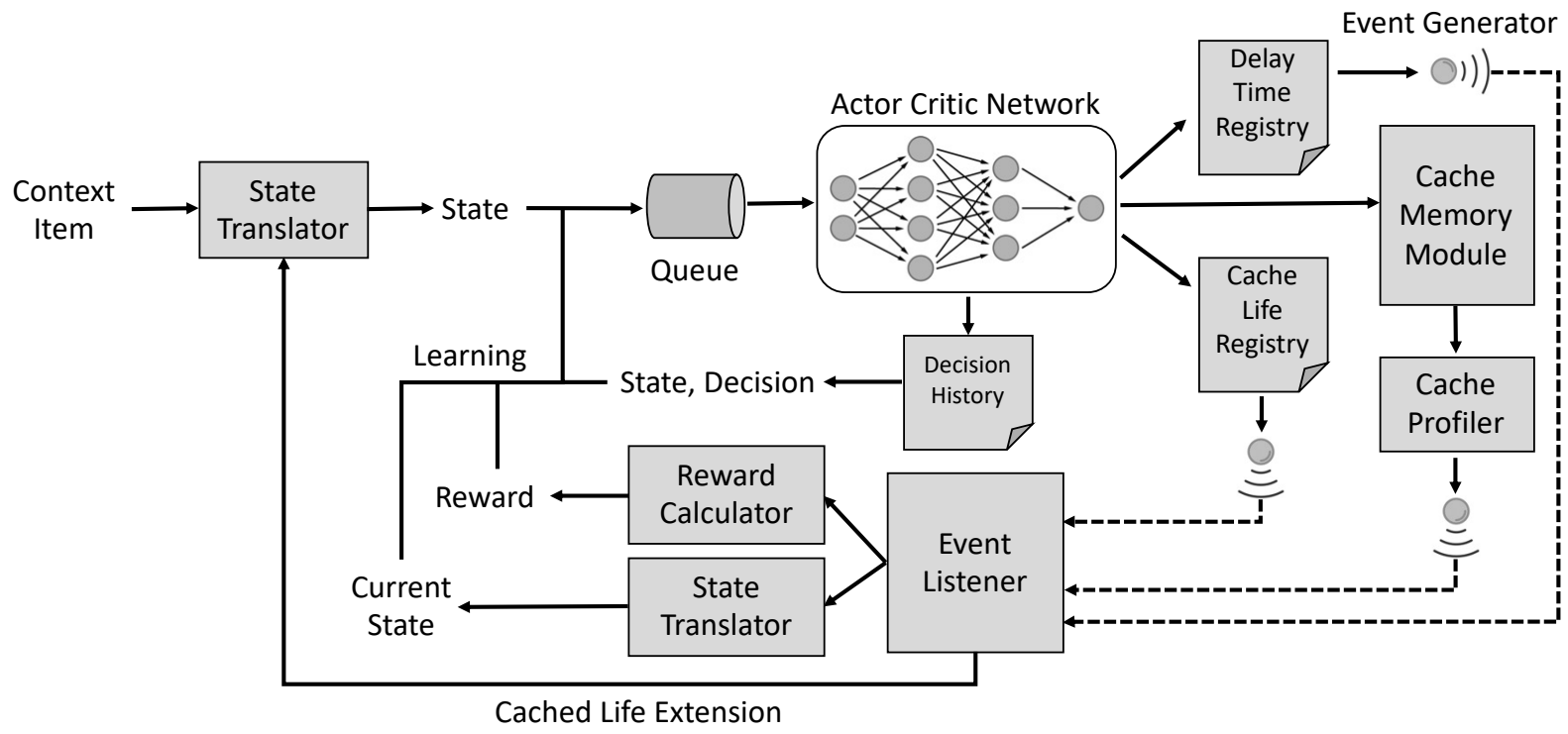

Figure 4: Process of the Actor-Critic Agent.

The function $f$ assigns a large negative reward to items that has not gained any or very few intermittent subsequent accesses once cached. Such items are cost-inefficient to cache since they incur *hold-up costs* – the cost of occupying the cache space for no return. Function $g$ assigns a positive reward for $a_i = 0$ decisions for these items. But $|g| < |f|$ because we encourage the agent to make caching decisions ($a_i = 1$) considering our objective to minimize RT of time-critical context queries.

### 3.2 Deep Deterministic Policy Gradient Agent

The seconds strategy involves using a *policy search* method. In the previous strategy, *ACAgn*, the goal state is reached by attempting to estimate either of value ($V^{\pi}$) or Q-function ($Q^{\pi}$) for a given policy, e.g., the optimal policy ($\pi^*$). Given the lack of prior knowledge, e.g., the transitional probability distribution for each context-cache action pair, it is theoretically hard to accurately estimate at least one of these functions. Previous work in this respect suggests methods to parameterizing the policy ($\pi_\theta$) that could be tuned in search of the optimal policy based on the gradients of performance metrics, e.g.., if the performance metric is cost, the policy search follows that which minimize the gradient of the cost variation.

Policy search methods in general are faster to converge among RL based methods. In Section 2, we identified fast convergence as one of the main objectives of ACOCA and our previous work [38] in order to primarily, minimize the negative impact of: (i) the cold start problem, (ii) temporally varying characteristics of the query load, i.e., composition of context queries, and (iii) emergence of new context and queries, on the objective function. Therefore, in comparison to the *ACAgn*, an agent using a policy search method such as the Deep Deterministic Policy Gradient Agent (*DDPGAgn*) could converge to a long-term optimal policy avoiding local optimum at a lower convergence time.

We implement an actor-critic network which optimizes for the policies using the DDPG method. *DDPGAgn* follows the same process indicated in Figure 7 and calculation of the reward functions in (4), (5), and (6). We implement four deep neural networks – a Q-Network and a Target Network each for the actor and the critic to improve self-learning [23,26].

Consider the previous agents – *StatAgn*, and *ACAgn*. The "when" decision was defined based on the caching or not caching decision using it as a confidence metric, e.g., consider (3). However, the operational decision is either caching for at least the estimated period $CL_i$ or delaying till the "correct" time for $DT_i$, either of which is a continuous time value. We define a continuous action space $\mathcal{A} \coloneqq [-long, long]$ where $a_i \in \mathbb{R}$ for the *DDPGAgn* as resolve. Then the discrete actions of cache or not cache can be defined as follows:

$$discreteAction = \begin{cases} cache, CL_i = a_i; if\ a_i > 0 \\ not\ cache, DT_i = \frac{a_i}{W}; otherwise \end{cases} \quad (7)$$

Based on the work of Lillicarp [23] and using (9) and (10), when the optimal policy is deterministic ($\varrho: \mathcal{S} \leftarrow \mathcal{A}$), then (10) can be modified as follows to avoid the inner expectation:

$$Q^{\varrho}(s,a) = \sum_{s'} p(s'|s,a)[r + \gamma \sum_{a'}(\pi(a'|s') \times Q^{\varrho}(s',a'))] \quad (8)$$

Accordingly, $Q^{\varrho}$ can be learnt off-policy, i.e., using Q-Learning, using a transitional probability distribution generated by a different process. The actor uses a parameterized function $\varrho(s|\theta^{\varrho})$ where $\theta$ are the parameters. The critic implements the Bellman equation above in (8).

Another difference of the *DDPGAgn* to the *ACAgn* is the learning process. In *ACAgn*, we learn for each context item $i$ in the DH. This is expensive and the process does not generalize for heterogenous context items (which is a feature of context query loads) since the model attempts to optimize for each individual context item. It is a drawback when the context space is large. We use a mini-batch processing technique considering a recent decision sample of size $\vartheta$, e.g., 16 decisions, in each learning cycle to reduce the bias to skewed or corner cases. Further, we extend the concept of "soft" target updates in [23] given by:

$$\theta' = \tau\theta + (1-\tau)\theta' \quad (9)$$

The rationale is to stabilize learning by converging slowly to the optimal policy [23]. This strategy has two drawbacks: (i) long convergence time, and (ii) high processing resource utilization (which therefore disqualifies from being useful in limited resource nodes such as a router) but can produce a better long-term efficient solution, as we show later.

### 3.3 Comparing the selective context caching agents

In Table 4, we rank our three selective context caching agents theoretically on the basis of several key performance indicators. Rank 1 indicates the best performance whilst 3 suggests the worst performance among the three agents. Based on our objectives stated above, we opt to maximize performance efficiency while minimizing cost. The hypothesis below are derived using this table.

Table 4: Comparison between the expected performance of the agents

| Parameter | StatAgn | ACAgn | DDPGAgn |
|---|---|---|---|
| Convergence Time | 1 | 2 | 3 |
| Processing Time | 1 | 2 | 3 |
| Processing Expense | 1 | 2 | 3 |
| Short Term Optimality | 1 | 2 | 3 |
| Long Term Optimality | 3 | 2 | 1 |

Accordingly, we make the following hypothesis:

- The two RL agents will not converge to short term optimality like the *StatAgn*.
- The RL agents yield a negative return compared to the positive returns shown by the *StatAgn*.
- Performance of the *StatAgn* in terms of cost efficiency will reduce over recursive planning periods whilst the RL agents will improve.
- *DDPGAgn* performs better than the *ACAgn* in terms of cost efficiency in the long run.

In this section, we have described the two RL based selective context caching agents. In the next section, we will discuss the adaptive context caching process which incorporates them in further detail.

## 4 ADAPTIVE CONTEXT CACHING ALGORITHM

In this section, the adaptive context caching algorithm is discussed in detail focusing on the process flow, heuristics, eviction policies, scaling, and cache memory.

Figure 5 depicts the state-transition diagram of a context item. The ghost state refers to being moved to a buffer memory temporarily to evict until any substantive queries are served before freshness is completely lost. Figure 6 depicts the summarized flow chart for responding to context queries. It should be noted that an item is asynchronously evaluated for caching either when, (i) not cached, or (ii) $CL_i$ has elapsed. We define a *complete hit* as when a derived context available in the cache is complying with the freshness requirement (from the SLA). A *partial hit* occurs when a derived high-level context needs to be re-produced and refreshed using lower-level contexts items where at least one of them is fresh and cached. A complete miss is when neither of the above criteria are met.

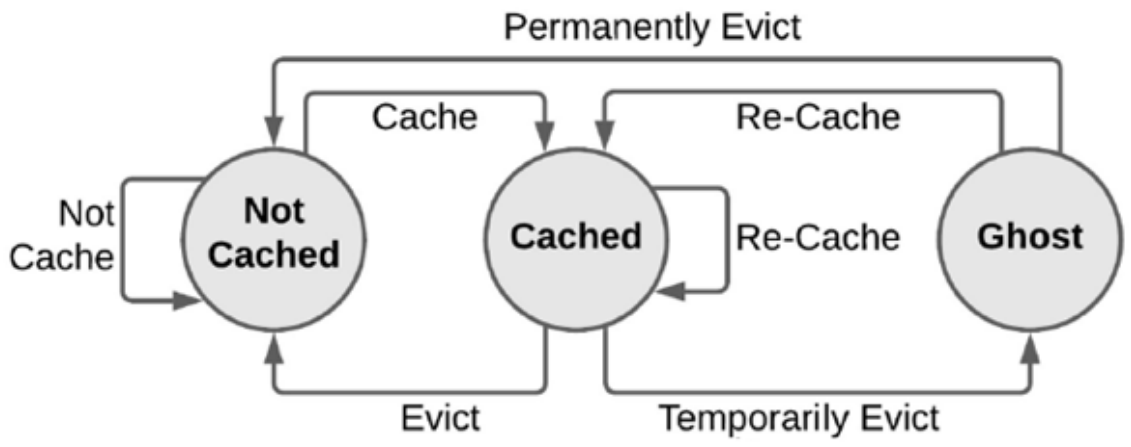

Figure 5: State transition diagram of a context item.

According to [24,38], context retrieval from provides accounts for 20-25% of the costs incurred by a context management system whereas the rest is on penalties, using the redirector mode. We indicate this in Section 5. Our cost minimization objective can be achieved primarily by minimizing the response latency when responding to context queries. Secondarily, cost can be minimized by managing the number of retrieval operations including for refreshing. We incorporate three solutions in this respect. Firstly, we introduce the criteria *is_spike* to bypass delaying evaluation until $DT_i$ elapses in order to handle bursts of queries that are typically a result of an event or an identified situation. Retrievals are minimized as a consequence of reduced processing. Secondly, the algorithm is optimized to minimize resource consumption and one of the strategies involves *sharing context retrievals* - multiple parallel queries sharing a single request-response pair per provider. All cached context items, e.g., context attributes, originating from the same context provider are refreshed from a single retrieval operation in this manner. Thirdly, the reactive refreshing strategy [24,38] is used in this paper for refreshing operations so that the retrieval occur only when it's necessary.

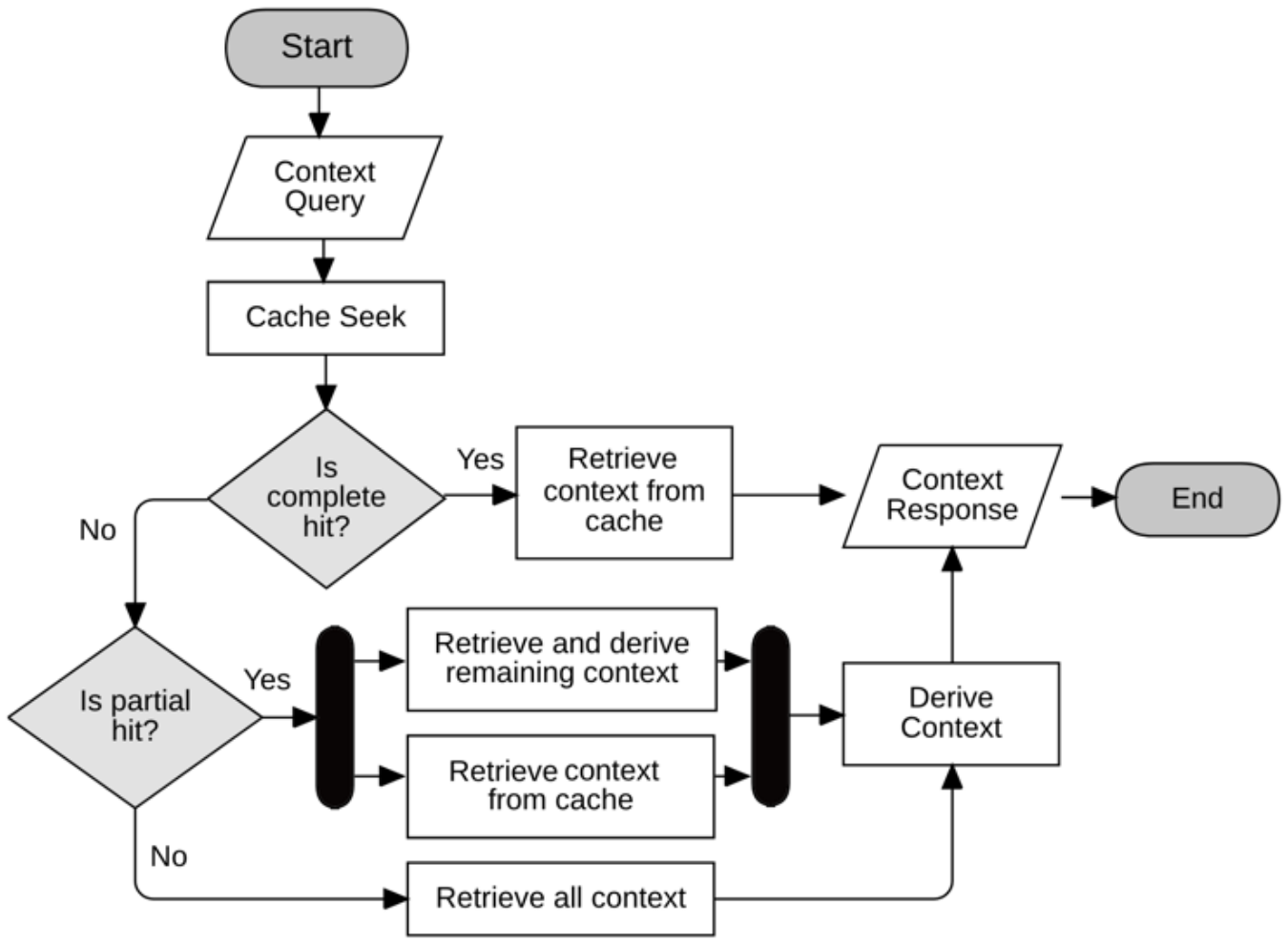

Figure 6: Summarized flow chart for responding to context queries.

### 4.1 Statistical Heuristics

One of the non-trivial problems in adaptive context caching is the lack of prior knowledge due to (a) emergence of new context information (unavailability of previously observed statistics), (b) lack of observed data by volume, (c) uncertainty in data due to variability in the environment, e.g., E[AR] of $Q_4$ under adverse weather conditions versus a calm day between 7:30-8:30am, (d) imperfections [17,21], and/or (e) incompleteness, e.g., having no observed data on AR of context query $Q_5$ between 3:00-4:00am although there exist in abundance between 7:30-8:30am. Blasco and Gunduz [6] explicitly state this challenge taking popularity profiles and Sheng et al. [31] develop their strategy based on no prior information. We adopt several simple heuristic solutions based on contextual features to overcome several of these issues in this work which are explained in this sub-section.

Consider the application of (4) for a context item that lacks previously observed data to calculate the expected values for HR. We calculate the effective cache lifetime using the formula indicated below based on Figure 7 to estimate the $\mathbb{E}[HR]$. The residual lifetime [35] remaining to cache, $ResiL = (f_{arrive} - f_{thresh}) \times \overline{ReL}/(1 - f_{arrive})$ given $f_{arrive} = 1 - \overline{ReL}/L_i$. HR yields to 0 when $f_{arrive} \leq f_{thresh}$ because $ResiL \leq 0$. Otherwise, considering a minimum $\mathbb{E}[HR] = 0.5$ that could be translated into the criteria $ResiL \geq 2/\mathbb{E}[\lambda]$, we define:

$$\mathbb{E}[HR] = \begin{cases} 1; when\ L_i \to \infty \\ \frac{\mathbb{E}[ReL]\times\mathbb{E}[\lambda]\times\mathbb{E}[AR]}{(\mathbb{E}[ReL]\times\mathbb{E}[\lambda]\times\mathbb{E}[AR])+1}; otherwise \end{cases} \quad (10)$$

We find a similar notion to (10) in the form of probability of an access being a hit ($P_{hit}(i)$) in [33] for defining *GoodFetch* – a cached item that balances access and update (defined as *refreshing* in our scope) frequencies.

Another approach to overcome this problem is to explore the state space using the $\varepsilon$-greedy method. We randomly cache context items decided not to cache at a probability of $\varepsilon$. Although it is simple to implement, complete random exploration cannot reduce the convergence time to minimize the effects of the cold-start problem if the state space is considerably large, unless the randomly cached items yield a high return from responding to queries. We use the Most Frequently Used (MFU) strategy to cache context items greedily in this paper. The choice of this heuristic was made based on experimental observations comparing against Most Recently Used (MRU) and complete random strategies, the discussion of which is beyond the scope of this paper.

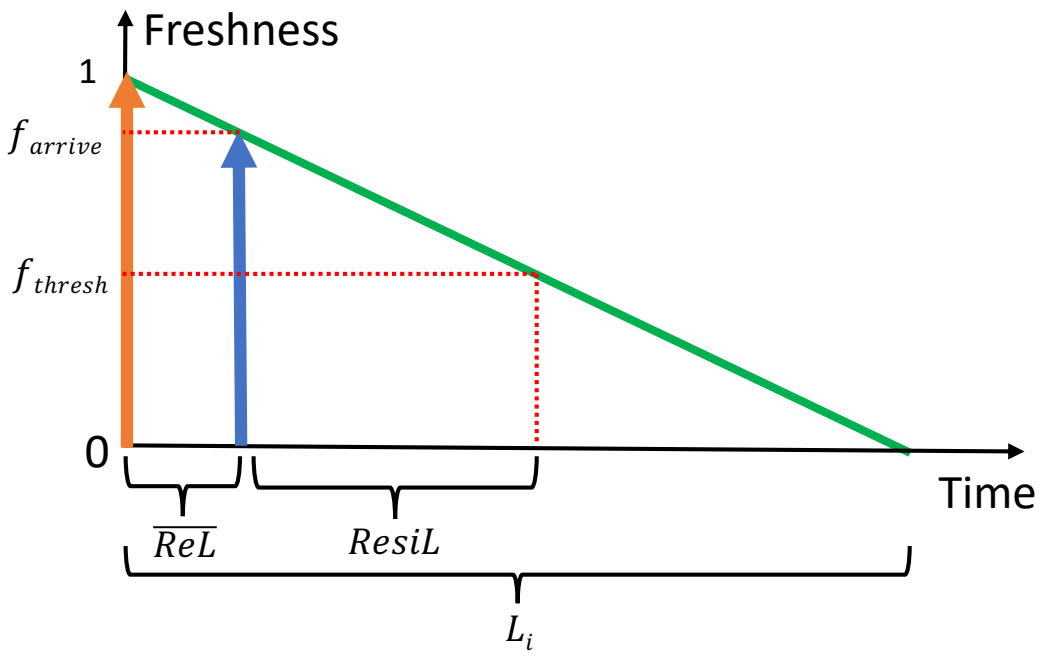

Figure 7: Estimating $\mathbb{E}[HR]$ using known parameters.

A further drawback of $\varepsilon$-greedy exploration in the literature is the lack of adaptiveness [18]. Context query loads are generally heterogeneous in nature but can often exhibit, (i) homogeneity, e.g., a higher proportion of queries searching for alternate routes after a roadside accident, and/or (ii) consistency, e.g., recurring spatial-temporal query patterns such as searching for car parks during rush hours. Constant $\varepsilon$ value could result in unnecessary explorative caching in these scenarios resulting in cache-inefficiencies, especially when the cache memory is limited. It is rather more optimal to exploit – relying on the known actions that generate higher reward, e.g.., caching item *i* which generated a positive reward from caching, when it was in the same or a similar state. On the contrary, more exploration, i.e.., large $\varepsilon$, could be beneficial when the queries are highly random. We introduce adaptiveness to $\varepsilon$ by modifying $\varepsilon = \varepsilon \pm \zeta$ after each $cycle_{learn}$ [23] as a result. The dynamic nature of the context queries is quantified by $\overline{Reward}$ for all the caching decisions since the last $\varepsilon$ modification event. For instance, consider a stream of queries that request a set of new context items. A trained SCA may decide not to cache due to the lack of statistical observations and negative expected values, e.g., (4) could yield a negative average reward despite the items exhibiting a high AR. $\varepsilon$ is incremented if $\overline{Reward} < Reward_{max}$ or decremented otherwise to adjust for this situation.

We also introduced an adaptive $\delta$ that modifies $\delta = \delta \pm \omega$ after a constant number of decisions referred as the *learning cycle* ($Cycle_{learn}$) to reflect the current level of volatility in the query load. For instance, $\delta$ is reduced by $\omega$ in default for each window unless a $cycle_{learn}$ occurred that increments $\delta$ by $\omega$. The rationale is $cycle_{learn}$ occurs frequently when the queries demand a heterogenous set of contexts to be evaluated for caching – a feature of volatility

### 4.2 Cache Memory

In-memory caches are faster yet expensive compared to persistence level caches used in literature. Further, devices that share memory units with cache, e.g., routers, suffer from data volatility. Although contemporary in-memory technologies such as Redis [46] support replication across distributed nodes, context cannot be recovered but only reproduced. This is due to the transient nature of context data that does not guarantee any recovered data to comply with quality of context parameters, e.g., freshness. For the purposes of this paper, we can assume that context originating from IoT devices are small and reliably retrieved that context can be reproduced fast (based on related literature [3]). So, we opt to do in-memory context caching based on the following reasons: (i) low cache seek time suitable for responding to time-critical queries, and (ii) to test the generalizability of ACOCA in a system where memory is shared with cache.

Medvedev et al. [24] proposed a logical cache hierarchy which is extended in this paper. Consider Figure 8 below. It depicts an alternate simplified context hierarchy. As we discuss in Section 4.3 and depicted in Figure 9, the context caching decision process follows this hierarchy either top-down, e.g., for eviction, or bottom-up, e.g., when performing context cache placement and refreshing. We identify several advantages of defining the unit of cache memory logically rather than physically. Firstly, it separates the complicated physical memory management tasks from the logical cache. For instance, consider a cloud-based cache memory where the data has physically fragmented across multiple distributed instances. The defragmentation routine may result in scaling down the number of physical instances. But the logical cache would be oblivious to this situation. Secondly, it supports maintaining the logical affinity of hierarchically related context items. Therefore, the cache memory is scaled in logical units; however, corresponding to the context item that is highest in the logical hierarchy, e.g., in terms of its entity units in low-level context cache.

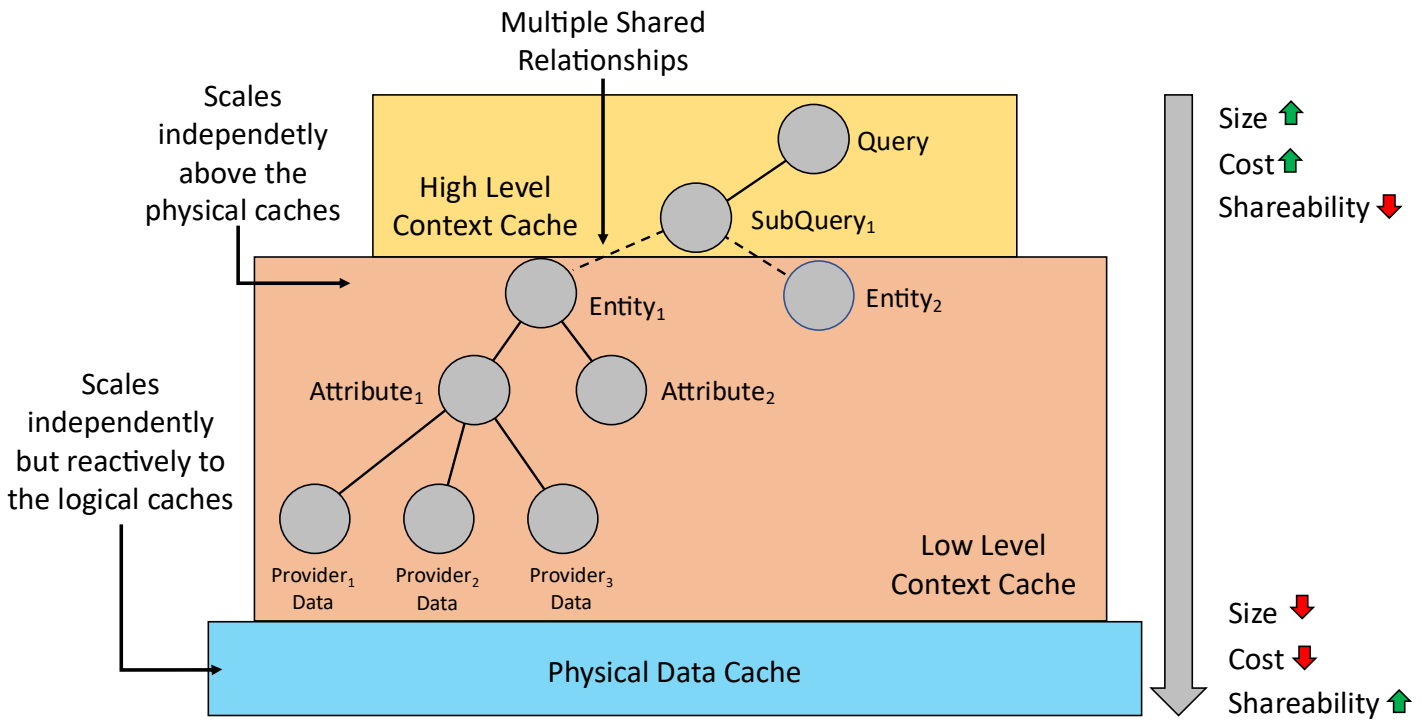

Figure 8: Simplified context cache hierarchy [37].

### 4.3 Eviction Heuristics

One of the features of caches in a distributed system is the diversity of capacity and technology. Devices towards the edge are often limited in size and shared whereas dedicated scalable cache memories could be found in the Cloud. Critical evaluation of existing literature suggests the advantage of adopting scalable caches to cost-effectively handle dynamic loads. Although comprehensive solutions for issues such as post-scaling degradation or performance cliffs [11,34], and distributed fragmentation [46] are yet to be investigated, overall performance gains outweigh the disadvantages. We discuss a criterion to adaptively scale context cache memory as a part of the adaptive context caching and eviction processes in this section based on this rationale.

We assume that the pay-as-you-go scheme of cloud caches involves making an upfront payment for each unit of cache, i.e., 2GB. The cost of caching is distributed among all the items that share the cache memory. So, an underutilized cloud cache is cost inefficient. According to Figure 9, we attempt to evict an item(s) only when the remaining cache space is insufficient. Considering the logical hierarchy of the context cache, a top-down approach to eviction is proposed in order to release a large amount of space in a single eviction cycle. We define two types of evictions: (i) mandatory, and (ii) selective. Mandatory eviction involves removing a context item, i.e., entity, and all its context data lower in the logical hierarchy, e.g., attributes from cache. Selective eviction removes only lower-level contexts which satisfy the decision criteria of the eviction algorithm. For example, only the *speed* attribute is evicted of the entity car, while the attribute *direction* remains in the cache. We use a threshold ($\eta$) based approach to categorize the types of evictions, the specific definition of which depends on the eviction algorithm.

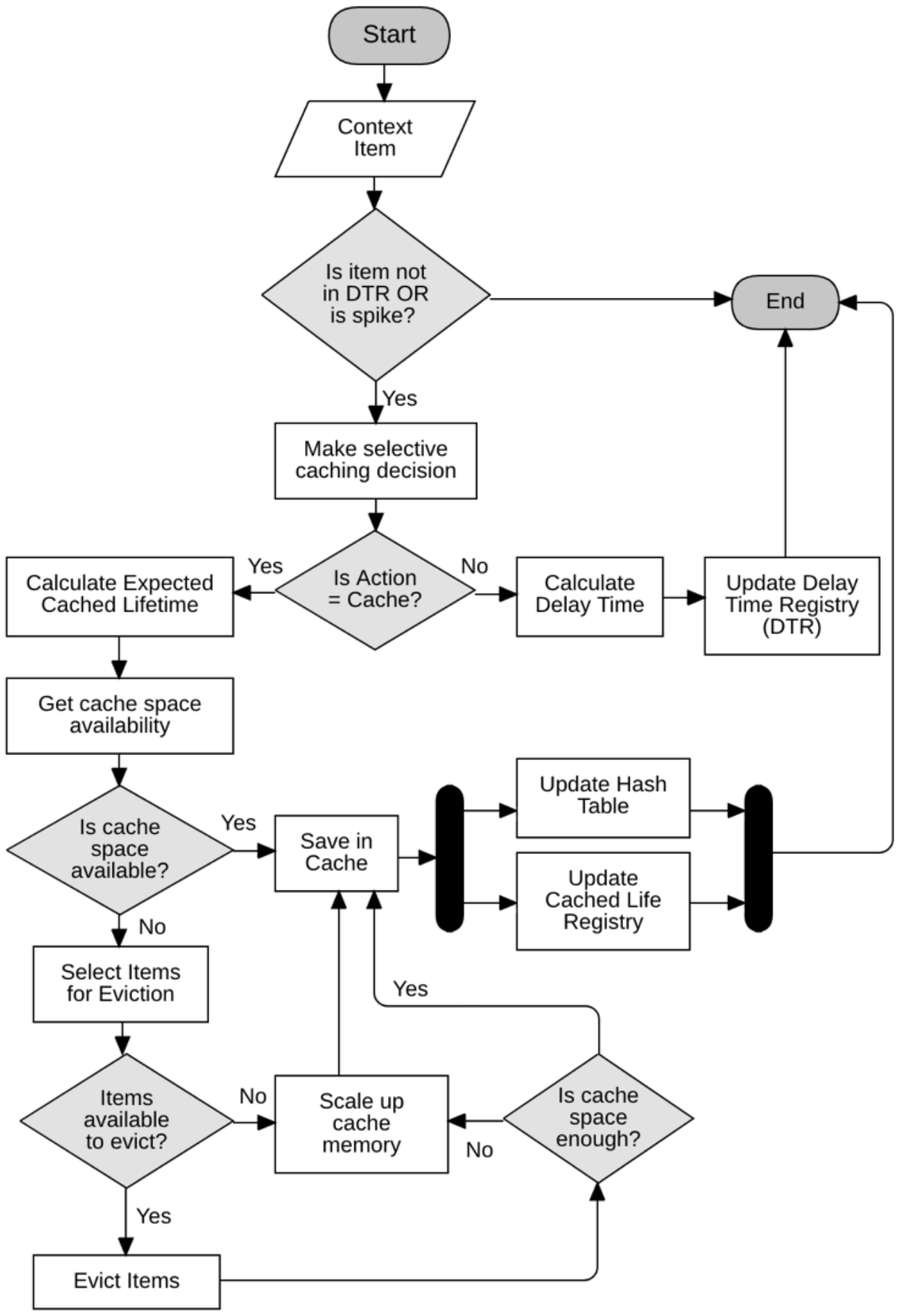

Figure 9: Simplified context cache hierarchy.

We compare the performance of windowed- (i) random selection of item to evict, (ii) Least Frequently Used (LFU), (iii) Least Valued First (LVF) [24], and (iv) our new Time-Aware Hierarchical (TAH) eviction policy that combines the advantages of time-awareness in TLRU [5] with LVF [2] and LFU as indicated in Figure 10 for low-level cache. Hierarchical aspects of TAH involves applying LVF and LFU as selection criteria for eviction at entity and attribute levels respectively. Previous work in caching state that access to objects follows a Zipf-like distribution [33,39] and popular items are expected to create higher hit rates [39]. Raw data from context providers are the most accessed in this regard since they are most useful to be reused and repurposed. We indicated in Figure11 that attributes can have multiple providers, which makes context attributes share this feature. Context attributes are the lowest in the logical context hierarchy. So, evicting based on frequency of access is logical. According to [14] as well, sub-queries are based on entities. We have also shown above in Section 2 that criticality differs for context queries comparing $Q_1$ and $Q_7$. It is more useful to evict entities based on their significance in contribution to queries which we quantify using "value". The TAH policy is therefore designed accordingly.

It is important to note that all eviction policies other than the TAH policy are not time-aware, i.e., an item could be evicted despite its estimated $CL_i$ has not expired. The value of a context item is calculated as follows where $\hat{I}$ refers to the most popular context item of the same logical level, $\ddot{I}$ refers to the context item that takes the longest latency to retrieve and $\Delta CL_i$ refers to the remaining cached lifetime:

$$\mathbb{E}[HR] = \begin{cases} 1; when\ L_i \to \infty \\ \frac{\mathbb{E}[ReL]\times\mathbb{E}[\lambda]\times\mathbb{E}[AR]}{(\mathbb{E}[ReL]\times\mathbb{E}[\lambda]\times\mathbb{E}[AR])+1}; otherwise \end{cases} \quad (11)$$

As an example, $Value_i$ is low and a possible candidate for eviction when an entity is relatively unpopular, $\Delta CL_i \to 0$ or $\Delta CL_i$ has elapsed and the context item can be retrieved fast so that reproduction is relatively easy.

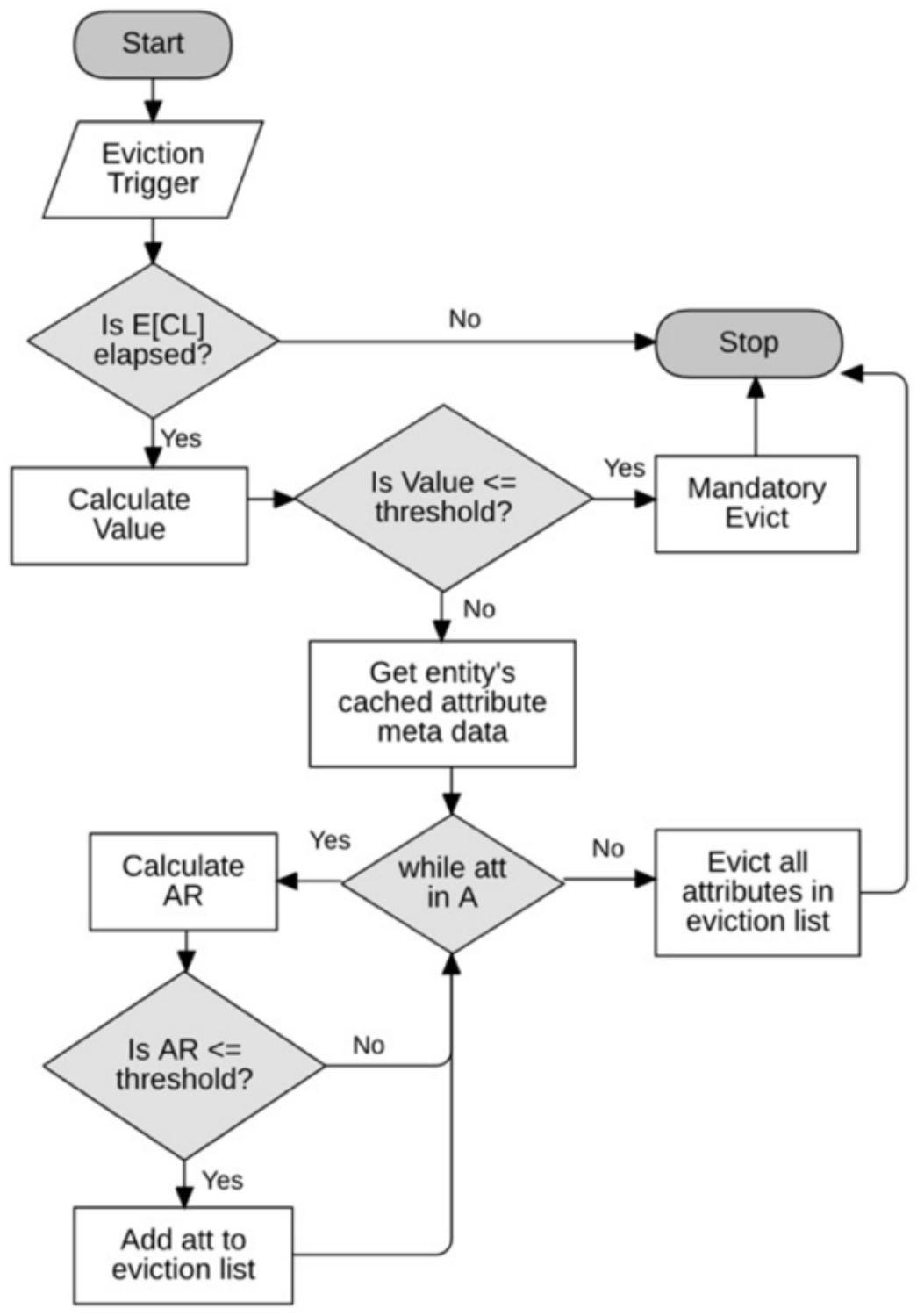

Figure 10: Time aware hierarchical eviction policy considering a context item.

It should be noted that items that avoid being evicted by the end of their $\mathbb{E}[CL_i]$ (set at the caching epoch) are reevaluated and the Cache Life Registry (CLR) updated with a recalculated $\mathbb{E}[CL_i]$. This extends the residence period of an item in the cache.

## 5 IMPLEMENTATION AND EVALUATION

In this section, our adaptive context caching algorithm is evaluated and the selective caching agents are compared. We evaluate the optimality of the caching decisions using the metrics defined below to establish our approach as a suitable cost-optimal solution for adaptive context caching.

### 5.1 Prototype and Experimental Setup

We developed our proposed methodology as a proof-of-concept (POC) using Python 3.8.2 mainly comprising of six components: (i) Context Query Execution Manager (CQEM), (ii) Storage Query Execution Agent (SQEA), (iii) Selective Caching Agent (SCA), (iv) Cache Operations Module (CaOpM), (v) Cache Memory Module (CMM), and (vi) Context Service Resolver (CSR). CQEM coordinates the query execution and performs data transformation operations while SQEA executes cache access operations. SCA makes the caching decision and the CaOpM performs context data assignment to cache, refreshing, and eviction operations. CMM is an independent entity that self-monitors, scales, and stores context for fast access whilst CSR selects relevant context providers and handles communication with context providers. Each of these components are multi-threaded, e.g., data from context providers for a query is retrieved parallelly and access concurrency to shared resources (i.e., hash tables) are controlled using hierarchical write-locks whereas shared services (i.e., SCA) incorporate the publisher-subscriber pattern that reads from First-in-First-out (FIFO) queues. We used GRPC for inter-service communication. This design strategy focuses on massive scalability in a distributed environment by creating autonomous units to simplify the complexity of the adaptive context caching problem. The architecture of the system is illustrated in Figure 08. All functionalities other than those in the primary execution path depicted in Figure 09 are executed asynchronously. Therefore, the response time of a query is independent of these processing times including making the selective caching decision.

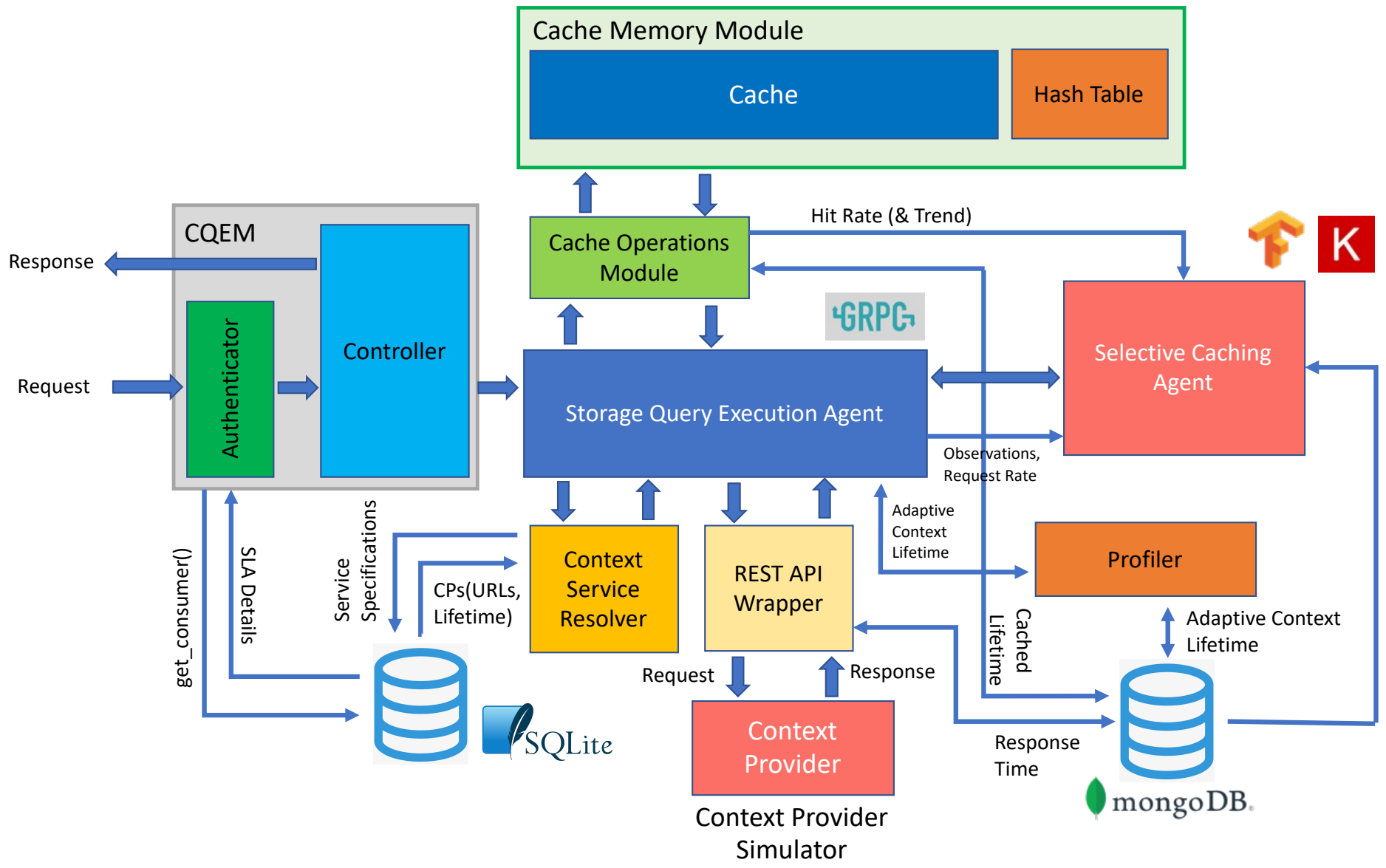

Figure 11: Architecture of the ACOCA-A prototype.

We used SQLite to handle structured data repositories such as the context services registry, service descriptions (SD) store, and resources lookup. MongoDB was used for maintaining logs, statistics, and historic context repository.

It is assumed that all context providers expose REST APIs, which we simulate using a proprietary IoT data simulator extended from our previous work [38]. Response latency of each context provider is normally distributed, e.g., for $GET : /cars?vin =$, $\overline{RL} = 0.997s$ and $\sigma^2 = 0.012$.

We consider a static context lifetime ($L_i$) and a linear freshness decay function (see Figure 11) to establish a controlled environment that focuses on our core objective, though assuming that $\mathbb{E}[L_i]$ for a given planning period (*PlnPrd*) is known using our previous work [38]. Therefore, $L_i = \max(1/SR_q, \mathbb{E}[L_i])$ for any transient context data originating from a context provider.

The *ACAgn* and *DDPGAgn* implements fully connected neural networks for the actor and critic networks using TensorFlow 2.0 and Keras. It contains two hidden layers containing 512 and 256 neurons respectively. We resorted to a small number of hidden layers in order to minimize the performance overhead so that the *StatAgn* could be compared more reliably. Rectified Linear Unit (ReLu) was used as activation function for hidden layers. SoftMax and Tanh was used for activation of the output layers of *ACAgn* and critic network of the *DDPGAgn* respectively. Mean squared error was used to calculate the loss between the Q-target and the Q-value at the actor network. We used the Adam optimizer for learning. *Scipy* is used for mathematical operations including extrapolation. We implement the replay buffer for the *DDPGAgn* of size = 100.

Our approach is benchmarked against: (i) redirector mode, (ii) limited sized cache memory, and (iii) database mode (no eviction policy). We later discuss our findings against the results of several related work as well. The default logical cache size of a cache memory unit is three entities, considering the highest number of entities accessed by a single query (i.e., $Q_4$ and $Q_7$) among those stated for the motivating scenario. So, a limited cache memory can accommodate only 37.5% of all cacheable entities in the scenario.

Our motivating scenario explained in Section 1 contains eight context queries and twelve sub-queries involving eight entities, seven context services, and twenty-five context providers. All queries are assumed to be pull-based. We used a single Unix-based edge server having 8GB of shared memory and 2.4GHz of CPU to respond to context queries originating from this contained area. Each context provider responds with a unique set of context attributes and can be uniquely characterized, e.g., $SR_q$. We defined eight context consumers each of whom has negotiated at least one SLA with the context management system. A context consumer can also be a provider, i.e., driver in $car_1$. The scenario provides a complex situation involving stationary and non-stationary entities where meta-data about the entity is derived from the SD, i.e., VIN of $car_1$, location of $carpark_2$. We evaluate our Proof-of-Concept using a set of 1471 unique sub-queries using Apache JMeter to generate the query load conforming to a Poisson distribution where $\lambda = 2.5/s$. A sub-query is either an aggregate function, situational function [14] or a sub-tree in the context query hierarchy (refer Figure 2). We present only the caching occurring in the low-level context cache (refer to Figure 08) in this paper. Each test case was executed at least three times per session (i.e., 4413 context sub-queries) over two independent sessions and averaged. The initial values of the hyperparameters are as shown in Table 5.

Table 5: Parameters and values used in experiments

| Parameter | Value | Parameter | Value |
|---|---|---|---|
| $\varepsilon, \varphi$ | 0.5 | $\varepsilon_{min}$ | 0.001 |
| $\gamma$ | 0.9 | $\varepsilon_{max}$ | 0.95 |
| $\alpha$ | 0.0001 | $\beta, \tau$ | 0.001 |
| $\kappa, \mu, \nu, \eta$ | 1.0 | $\omega, \zeta$ | $\pm$0.005 |
| $Reward_{min}$ | 1.0 | W | 5 seconds |

| Parameter | Value | Parameter | Value |
|---|---|---|---|
| $\vartheta$ | 16 decisions | $x$ | 10 decisions |
| *short* | 1 window | *mid* | 5 windows |
| *long* | 10 windows | $Cycle_{learn}$ | 20 decisions |

For simplicity, we assume cost of occupied cache space by an item $i$ is small ($Cost_{space,i} \to 0$) since the device memory is used as cache. The SLAs between the context consumer and the context management system are defined simplistically containing the price per valid response ($Price_{res}$), freshness threshold ($f_{thresh}$), cost of penalty for an invalid response ($Cost_{pen}$), and the maximum tolerated response latency ($RT_{max}$). The SLA with the context provider defined only the cost per retrieval ($Cost_{ret}$).We suggest interested readers to refer to our repository[1] for further details about the SLAs, SDs, entities, attributes used in the experiments.

For the purposes of this work, we define two metrics to evaluate the returns when responding to queries using an adaptive context caching strategy. Firstly, we calculate and store the pessimistic return (*PessiRet*) considering the most expensive and the least valuable SLAs applied within each window. The most expensive SLA defines the most stringent QoS requirements underpinned by the highest $RT_{max}$ and $Pen_{del}$. The least valuable SLA offers the least $Price_{res}$. These SLAs are selected at the end of each window by observation. *PessiRet* is calculated similar to (3) using observed actuals instead of expected values. Secondly, the total return from responding to all context queries during the observed period is calculated using $TotalRet = \sum_{\forall t} \sum_{\forall i} Ret(t)_i$.

In addition, we will consider the variation of response time, probability of delay (PD), throughput, cache throughput, hit rate (HR), number of context retrievals from providers, and the number of entity and attribute evictions for comparison.

### 5.2 Results and Discussion

In this sub-section, we present a summary of the observed results from the experiments and interpret the data. Errors indicated conform to a 95% confidence interval.

The following designations are used for graphs and figures in this section. The solid and dotted lines represent the use of scalable and limited-by-size cache memories respectively. Blue, yellow, green, and orange colours represent random, LFU, LVF, and our TAH eviction policies respectively. Red is used for the no eviction policy. A round marker is used to indicate explorative caching using the $\varepsilon$-greedy method and triangle markers when no exploration is performed.

First, we observed the impact of caching a context item on the overall performance of a sub-query. The response time was improved by ~85% compared to redirector mode. SCA was successful to minimize caching intermittently requested items irrespective of the caching agent. *ACAgn,* however, takes twice the time it takes for *StatAgn* to reach these conditions because of the cold-start problem which could be observed in the stepwise behaviour in Figure 12. Then we compared the overall performance of our methodology when limited and scalable cache memories are used. The summary is in TABLE 6.

Table 6: Comparison of agent performance using scalable and limited sized cache memories

| Cache Size | RT | HR | PD |
|---|---|---|---|
| No cache | $1.3843 \pm 0.01$ | $0 \pm 0.00$ | $0.6712 \pm 0.00$ |
| | | StatAgn | |
| Limited | $0.9903 \pm 0.09$ | $0.7031 \pm 0.04$ | $0.5197 \pm 0.05$ |
| Scalable | $0.6288 \pm 0.05$ | $0.9007 \pm 0.02$ | $0.4382 \pm 0.01$ |

| Cache Size | RT | HR | PD |
|---|---|---|---|
| | | ACAgn | |
| Limited | $1.2262 \pm 0.18$ | $0.4522 \pm 0.08$ | $0.6346 \pm 0.02$ |
| Scalable | $1.1719 \pm 0.17$ | $0.4471 \pm 0.07$ | $0.6077 \pm 0.05$ |
| | | DDPGAgn | |
| Limited | $1.1843 \pm 0.18$ | $0.6030 \pm 0.10$ | $0.6207 \pm 0.02$ |
| Scalable | $0.9728 \pm 0.15$ | $0.7283 \pm 0.09$ | $0.5526 \pm 0.03$ |

Using the *StatAgn*, we observe that that exists a strong negative correlation between RT and HR using a limited cache at -0.9672. The correlation coefficient was -0.7621 otherwise. There exists no correlation between RT and PD using a limited cache memory although there is a 0.8531 strong positive correlation in a scalable environment. Considering corresponding variances of 1.098 and 0.694, we can derive the reason for the above observation attributing to the higher number of cache evictions owing to constant variations in RT. The same feature of variance can be observed with other metrics as well. Therefore, limited-sized caches do not guarantee a consistent QoS compared to a scalable one.

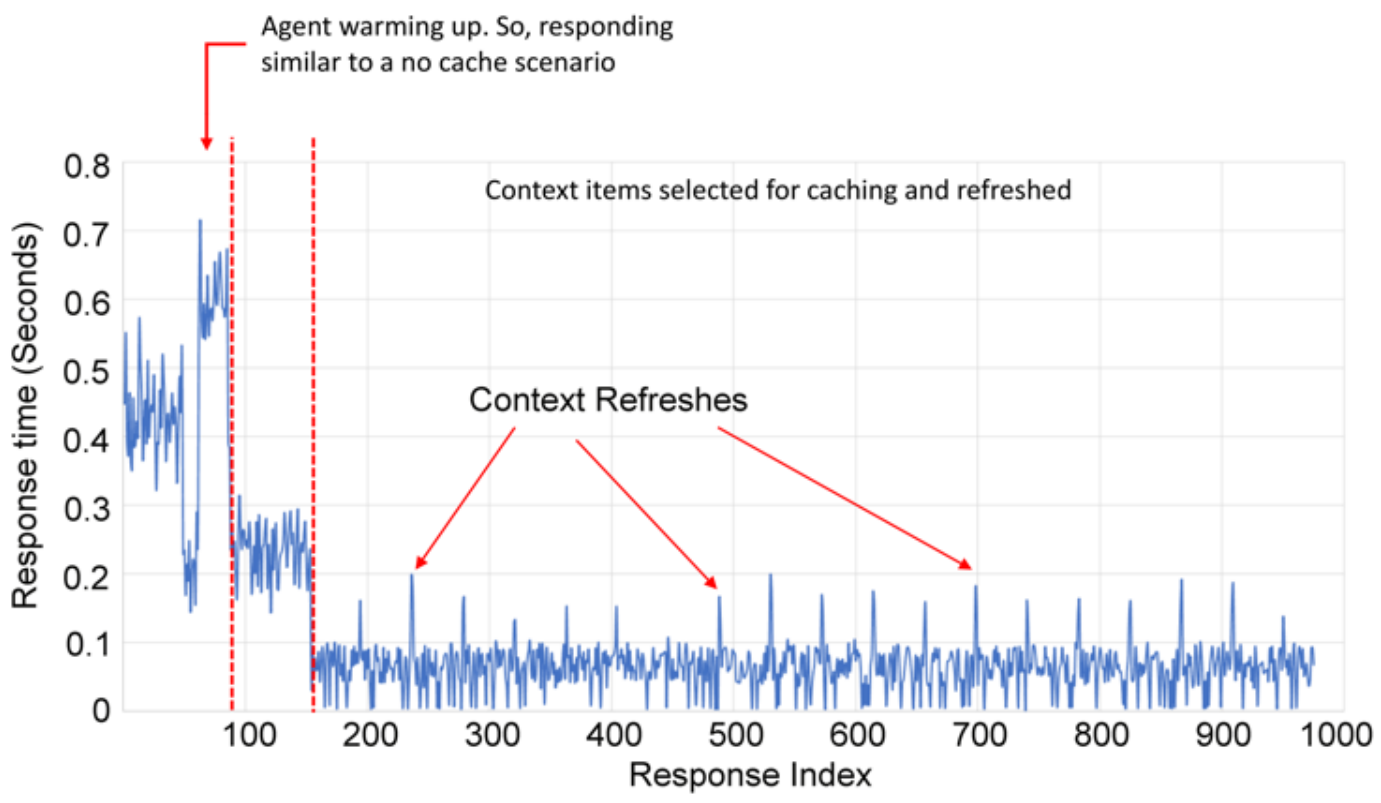

Figure 12: RT of a sampled sub-query per and post caching using *ACAgn*.

The *ACAgn* and *DDPGAgn* show strong correlation between RT-HR and also RT-PD irrespective of cache size. We summarize those results in Table 7 below. It conforms to our assumption of $P_{DC,i} \propto 1 - HR(T)_i$ for (4).

Table 7: Correlations of RT, HR, and PD for RL based agents

| Cache Size | RT-HR | RT-PD | HR-PD |
|---|---|---|---|
| | | ACAgn | |
| Limited | -0.7598 | 0.8107 | -0.5895 |
| Scalable | -0.7710 | 0.9094 | -0.7088 |
| | | DDPGAgn | |
| Limited | -0.8639 | 0.8349 | -0.6888 |
| Scalable | -0.8650 | 0.9322 | -0.8144 |

In the next two subsections, we will illustrate and discuss the results obtained for the performance efficiency metrics – $\overline{HR}$, $\overline{RT}$, throughput, and $\overline{PessiRet}$ of each selective caching agent under different configurations in the short-term. Then, we will compare the overall cost-efficiency. Efficiency of our strategy in the long-term is discussed next comparing the SCAs. Finally, we summarize our key observations from the experiments highlighting the key principle of ACOCA.

### *5.2.1 Actor Critic Agent*

First, the *ACAgn* was tested. We made similar observations on the consistency in QoS offering, correlations among RT-HR and RT-PD as the *StatAgn* in [37]. But we paid significant attention to exploratory selective caching using the adaptive $\varepsilon$-greedy policy given $\eta = 1.0$. Figure13 summarizes these results.

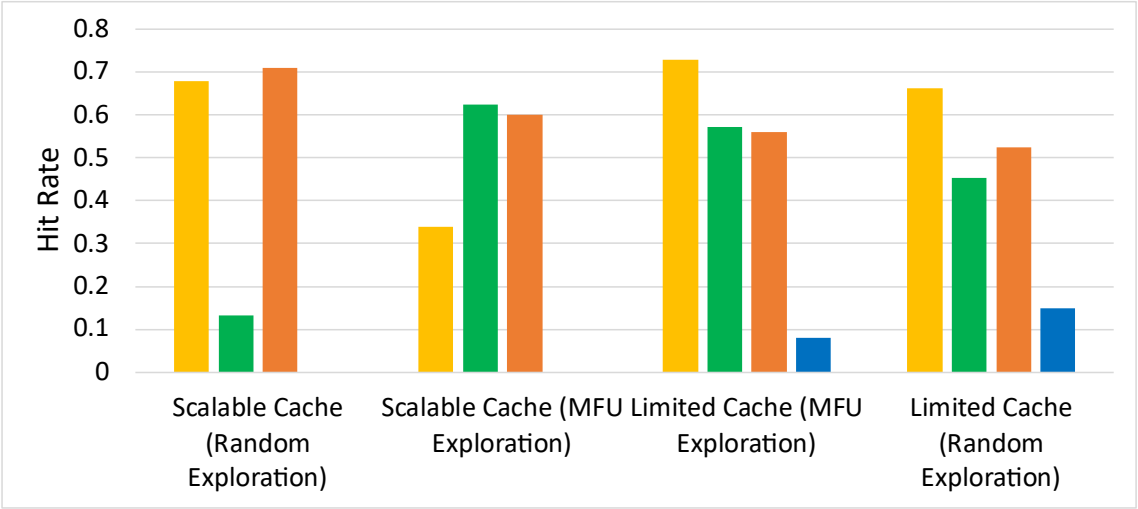

(a)

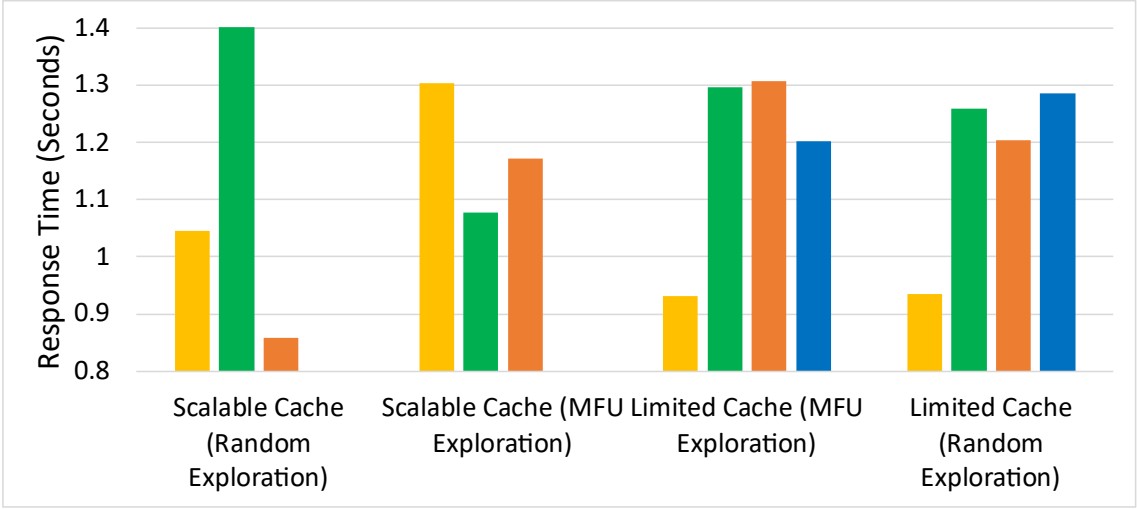

(b)

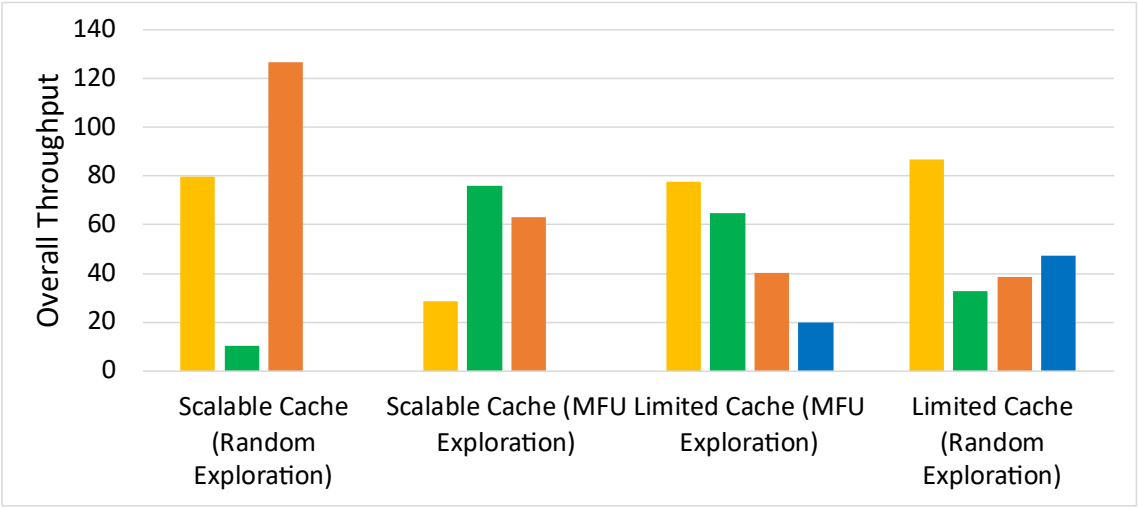

(c)

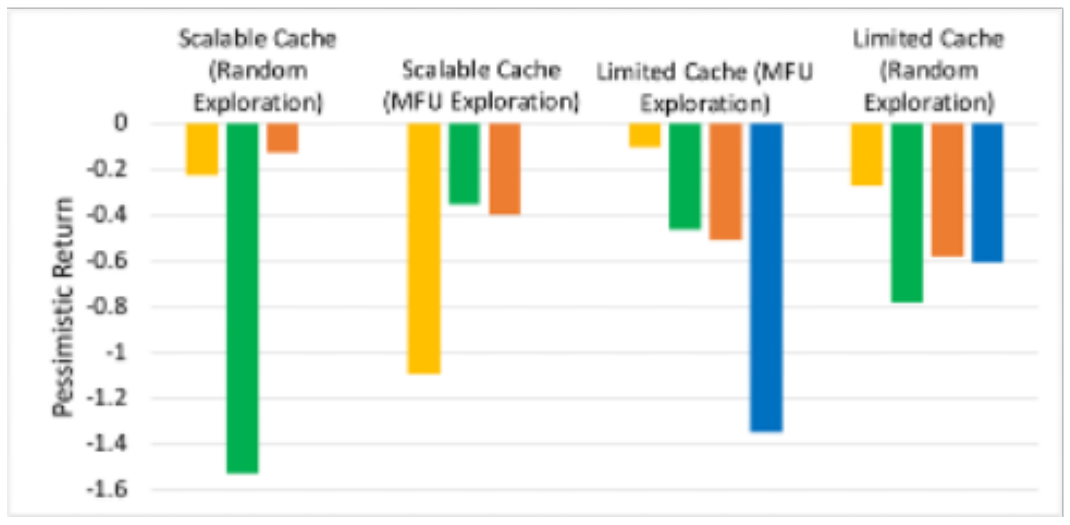

(d)

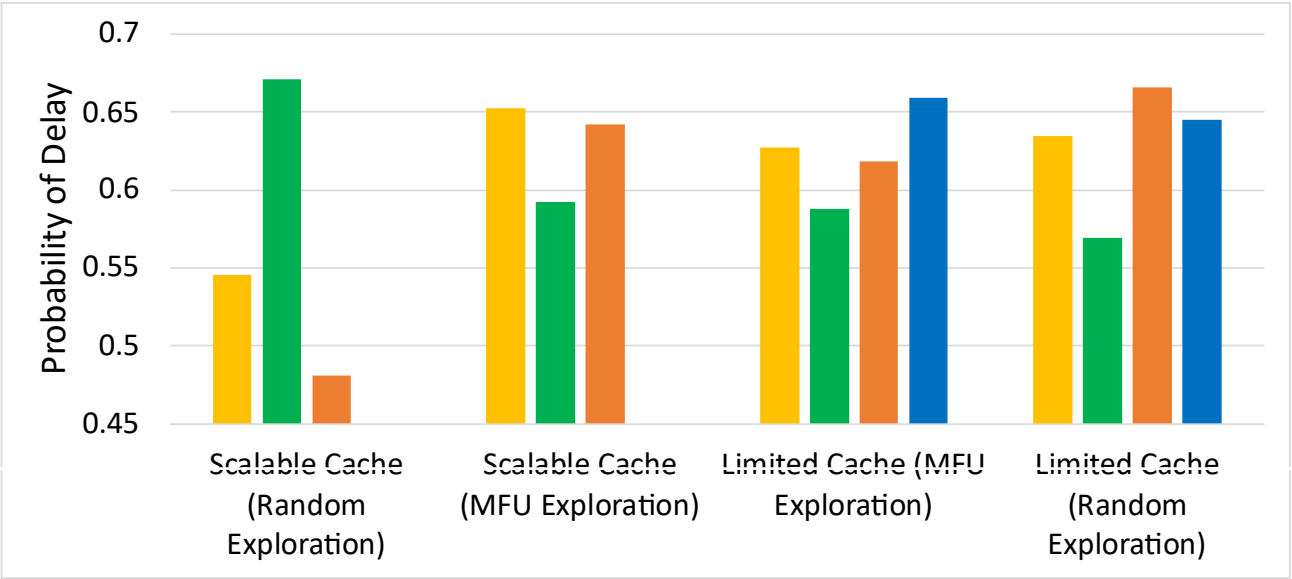

(e)

Figure 13: Performance metrics – (a) HR, (b) RT, (c) throughput, (d) PessiRet, and (e) PD of caching using *ACAgn*.

The agent makes random decisions when the transition probability distribution is non-existent and hence, we refer to this default as random exploration in this paper. There are several key aspects that could be identified. Firstly, the impact of explorative caching on context caching is only marginal given a limited-sized cache. This is because of frequent evictions. Secondly, only the LVF policy makes significant performance improvement on caching with exploration. This further confirms the rationale behind the design of our novel eviction policy to combine LVF with LFU. The novel policy shows significantly low PD, RT, a higher throughput, and the lowest negative $\overline{PessRet}$ among all the other results using a scalable cache memory given no explorative caching is done. The policy could be observed to be impacted negatively by exploration because the dominant feature in cached context item management is now the frequency of use, i.e., LFU of the policy and MFU of exploration. It shows more qualities in performance metrics using the LFU policy rather than LVF as it was observed with the *StatAgn* [37]. Thirdly, $\overline{PessiRet} < 0$ using scalable caches as well. We observed positive values using *StatAgn* under the same conditions [37]. In addition, they are worse than $\overline{PessiRet}$ compared to using a limited-sized cache memory. The reason could be observed in the minimized $PessiRet$ using explorative caching. This suggests that the RL model requires further learning to reach an optimal solution using exploration. Figure 14 illustrates this improvement over recurrently learning over the same *PlnPrd.*

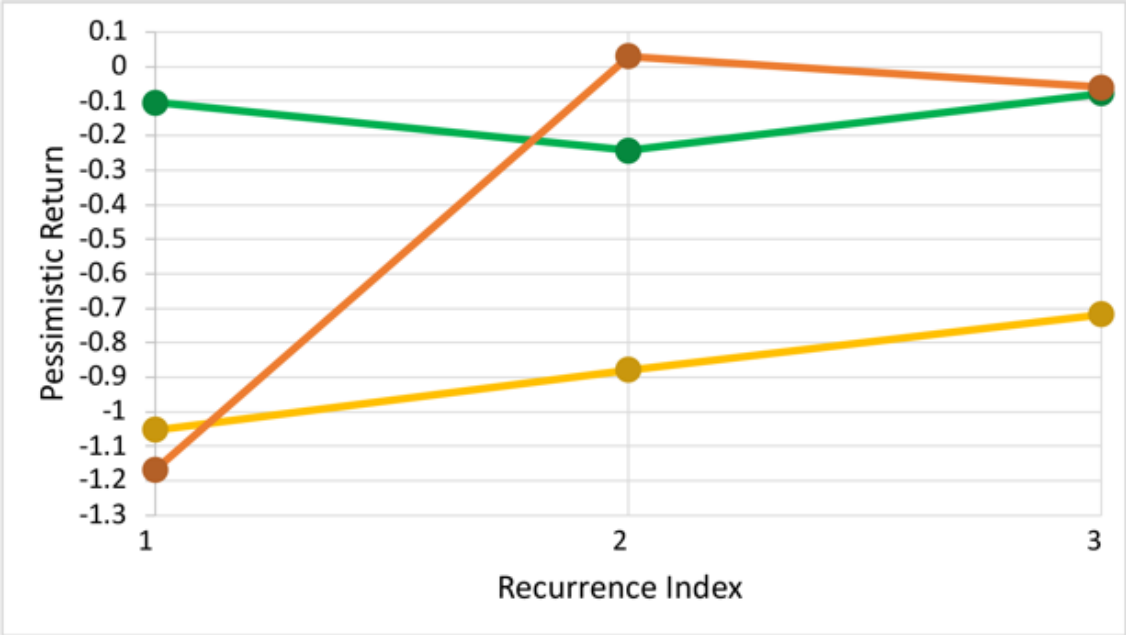

Figure 14: Improvement in *PessiRet* over three recurrent PlnPrds for *ACAgn*.

Considering that the query loads are generally heterogeneous and the objective to minimize the time to converge to an optimal solution, exploration is inevitable. LVF also reflects the features of our novel policy during the best-case scenario of a homogenous query load leading to no or low exploration. Therefore, the LVF policy could be identified to complement the *ACAgn* to deliver a cost-optimal context query management using a cache memory.

### *5.2.2Deep Deterministic Policy Gradient Agent*

Fourthly, the *DDPGAgn* was tested using similar setting to the *ACAgn* and results are summarized in Figure 15. The striking difference in results between the two RL based agents proves our rationale to parameterize the features for making the selective caching decision. LFU could be seen to perform consistently better compared to other policies with *DDPGAgn* including $\overline{PessiRet} > 0$ using scalable cache memory. The most important result is the low HR using scalable cache compared to high HR using limited cache. As we may see later in this section on long-term convergence, the optimal this random query load using scalable cache memory is biased towards redirector mode (i.e., not caching a many items). We can identify several reasons for this in this our setup. Firstly, the query load is random enough that caching an item result in higher probability for partial misses. Given we assume that cost of cache memory is ignorable, marginal return from of caching an item is minimized with size. The results suggests that the penalty incurred as a result of cache seeking and then retrieving or refreshing for a partial hit is costlier, that minimum number of items are cached. We observe this explicitly as the algorithm scaled the cache memory only up to 6 entities although in comparison, the *ACAgn* and *StatAgn* expanding up to 9 entities in size. Secondly, the high penalty is a result of high transiency of the context data. Thirdly, due to low marginality in cache memory cost, hold up cost is minimum (i.e., zero based on our assumption) resulting in the most cache worthy items being cached for longer periods. Using these same arguments for limited cache memory, the high HR can be justified. Consider the same items cached in scalable memory in a limited cache. Then the cache competition reduces partial misses resulting in higher HR. Overall, as we expected in our discussion in Section 3, *DDPAgn* has adapted well compared to *ACAgn* in all metrics.

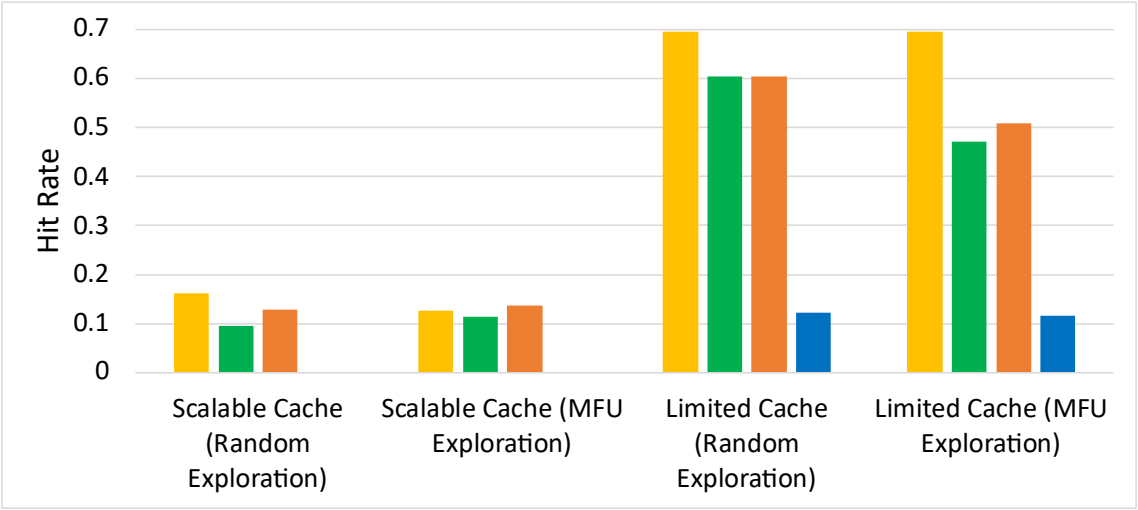

(a)

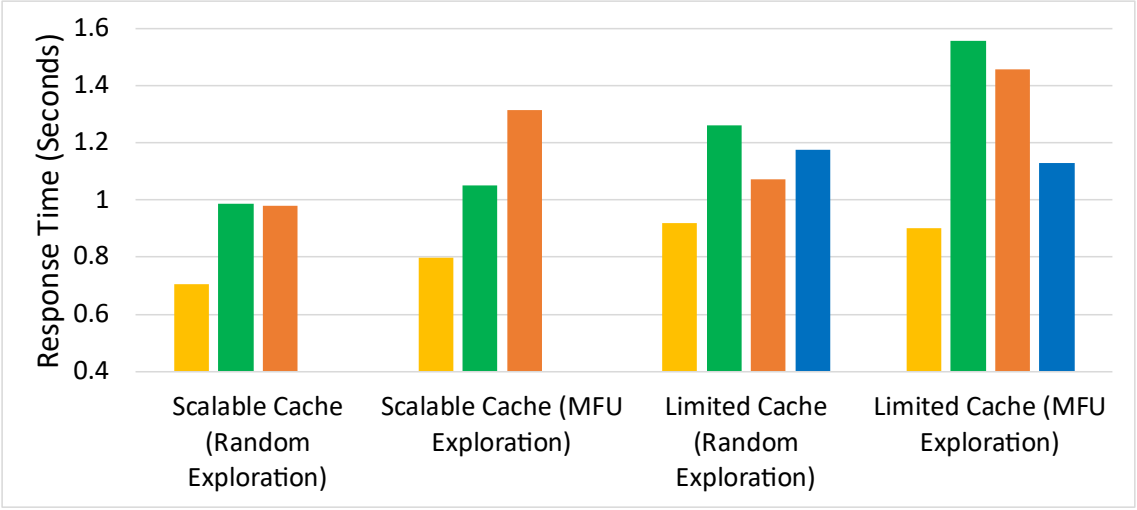

(b)

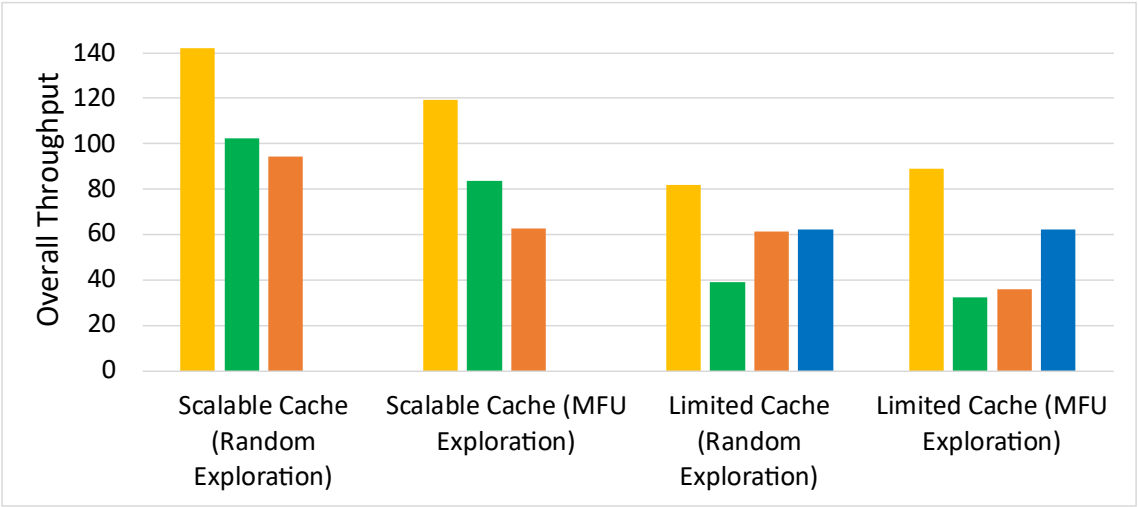

(c)

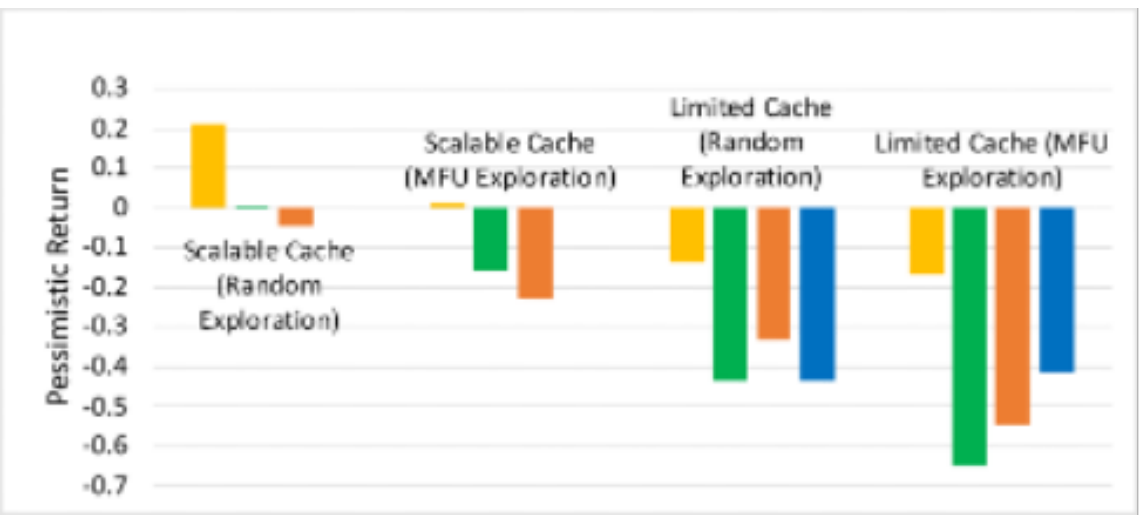

(d)

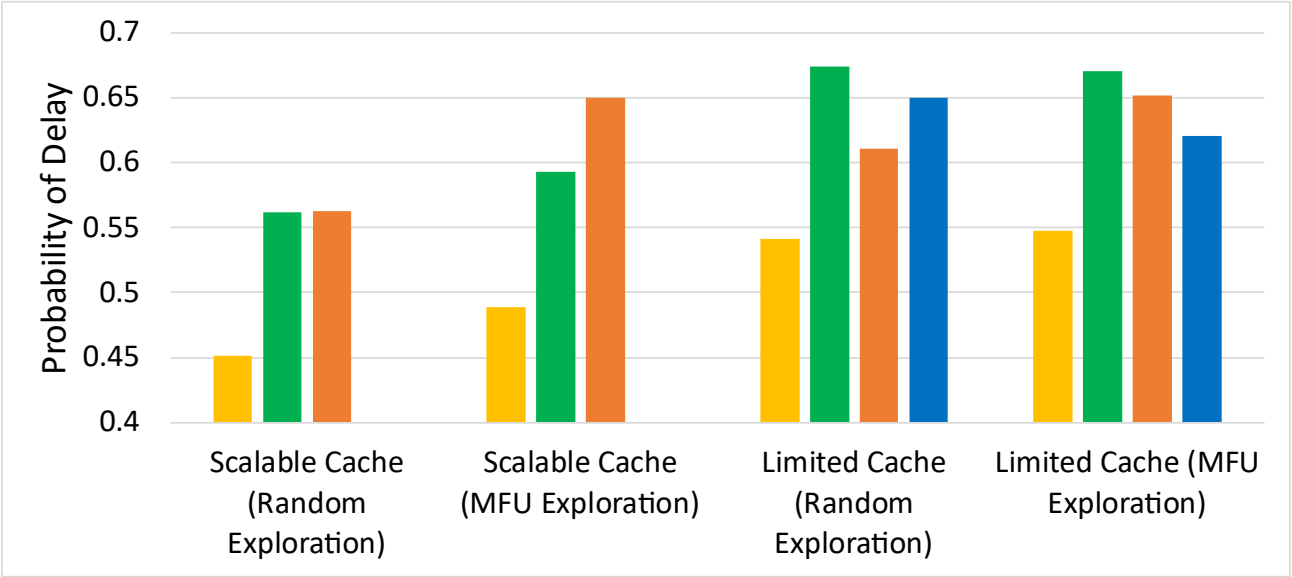

(e)

Figure 15: Performance metrics – (a) HR, (b) RT, (c) throughput, (d) PessiRet, and (e) PD of caching using *DDPGAgn*.

Figure 16 indicates the correlation between the number of entities evicted from the cache during the *PlnPrd* against the variance of $\overline{PessiRet}$ ($Var(\overline{PessiRet})$). There is a strong positive relationship among the two variables using a scalable cache memory using *StatAgn*, the relationship is weak for *ACAgn* and *DDPGAgn*. *StatAgn* is designed to cost optimize adaptive caching for the short term. Given the number of evictions is relatively constant against $Var(\overline{PessiRet})$ with a limited-sized cache suggests that the realized returns from responding to context queries vary only based on factors other than the selective caching decision. Hence, the caching decisions made by the *StatAgn* are short-term optimal and can be defined as the short-term goal state. In contrast, we can identify several reasons for the correlations of *ACAgn*: (i) the actor-critic network used in the *ACAgn* is configured to optimize for long term optimality where $\gamma = 0.9$, (ii) $\gamma$ and $Cycle_{learn}$ are defined statically in this work and the algorithm does not adjust to the variable dynamics of the environment in contrast to $\delta$ of its counterpart, and (iii) the correlation using the limited cache memory is a result of explorative caching since there are two distinct clusters which we believe would merge overtime when the optimal solution is reached on a theoretical basis.

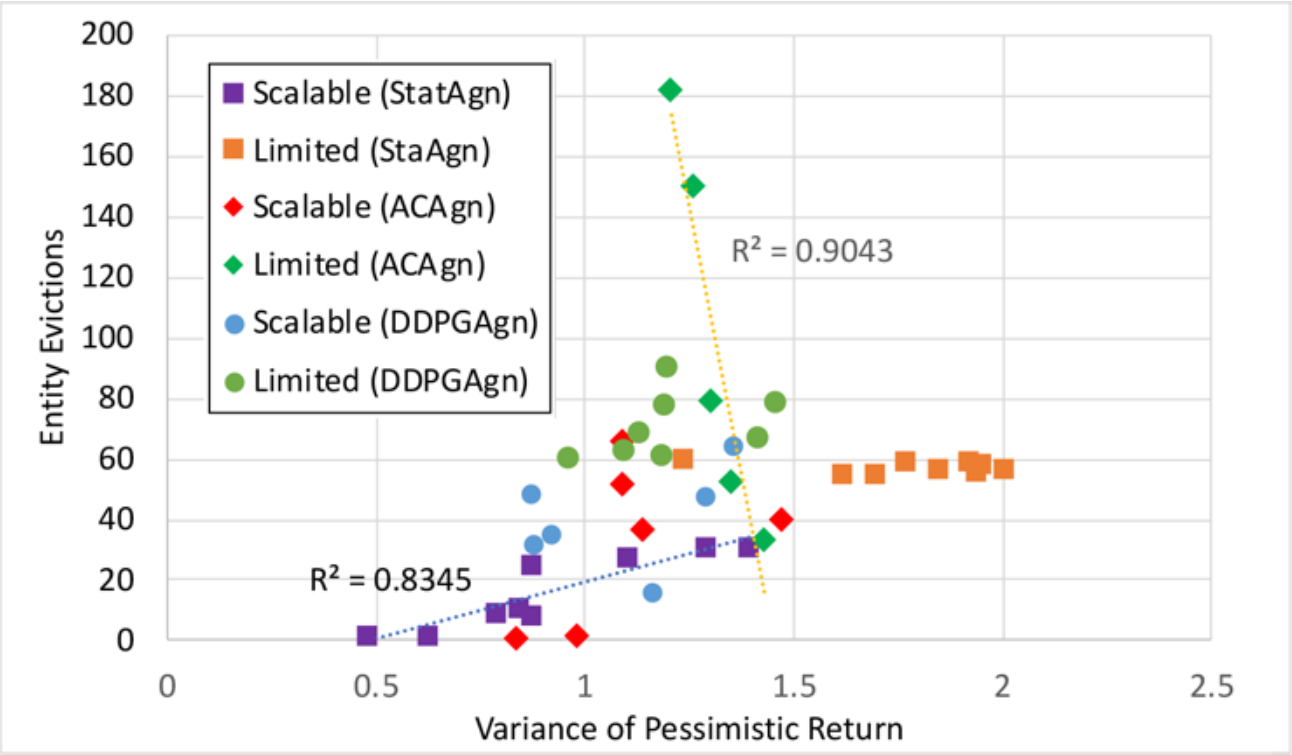

Figure 16: Correlation of entity evictions to variance in PessiRet.

Two important features of this convergence to short term optimality are: (i) convergence in a very short space of time, e.g., for less than 750 sub-queries and 5 minutes, and (ii) using limited amount of data. Similar work using RL have expended ~20,000 episodes [31], ~500 time frames [42], and more than 1000 time steps for an approximately 20% improvement in HR. So, our method extremely fast in comparison.

One of the important observations is that there exists no correlation between PD and HR for *StatAgn* despite any underlying policies. This points to our primary objective of cost optimization for the context management system. In Section 3, we discussed that it may not always be cost-optimal to cache all context items. Therefore, our approach has clearly been successful in achieving this.

### *5.2.3 Overall Cost-efficiency*

Next, the total returns of the context management system were calculated. Table 8 summarize this data using the *ACAgn* and the *DDPGAgn*. We have highlighted in bold the setting in which the maximum total return (lowest cost) is achieved in the tables. They match with the optimal configurations that were identified from above.

The total return using the Redirector mode (no caching) was -274.42 given the earning net of penalties and retrieval costs were -195.42 and 79 respectively. We showed in [37] that *StatAgn* for selective caching is clearly advantageous where a 47.94% and 34.57% increase in return was observed for the best setting for scalable and limited caches. Although the StatAgn was simplistic, it has exceeded the relative norm of achieved improvement at less computational expense.

The improvement in return is only 7.78% using *ACAgn* over a limited size cache memory. The rest of the settings for *ACAgn* yields a deteriorated total return. We can identify two main reasons: (i) the actor-critic model is not optimal and require further iterative learning, and (ii) the complex dynamics of context queries and data, i.e., transiency. The latter is a distinctive feature of context caching which we discussed elaboratively in Section 1 compared to data caching. So, due to the random and heuristic exploration of the state space for the purpose of learning, items that are cost-inefficient to cache are cached and vice versa until the $\pi^*$ is reached. In fact, results underlying a limited sized cache has always been worse than the random eviction policy. So, it confirms our hypothesis that the *ACAgn* is not converged in the short run.

Table 8: Total return for the context management system using ACAgn, and the DDPGAgn

| Policy | Exploration | ACAgn | | | DDPGAgn | | |
|---|---|---|---|---|---|---|---|
| | | Earning – Penalties | Retrievals Cost | Total Return | Earning – Penalties | Retrievals Cost | Total Return |
| Scalable Cache Memory | | | | | | | |
| No Eviction | Random | -157.63 | 63.17 | -220.80 | -116.95 | 56.50 | -173.45 |
| | MFU | -138.33 | 65.67 | -203.92 | -108.15 | 59.25 | -167.4 |
| LFU | Random | -146.97 | 86.33 | -233.30 | **-109.16** | **82.56** | **-191.72** |
| | MFU | -191.29 | 89.67 | -280.96 | -127.78 | 93.21 | -220.98 |
| LVF | Random | -193.95 | 80.75 | -274.70 | -157.14 | 93.31 | -250.45 |
| | MFU | -179.42 | 95.42 | -274.83 | -176.32 | 109.29 | -285.61 |
| Novel | Random | **-115.125** | **58.7** | **-173.82** | -160.88 | 70 | -230.88 |
| | MFU | -200.55 | 87.62 | -288.18 | -186.90 | 91.44 | -278.34 |
| Limited Sized Cache Memory | | | | | | | |
| Random | Random | -171.42 | 81.25 | -252.67 | -197.83 | 114.25 | -312.08 |
| | MFU | -190.97 | 106.21 | -297.18 | -184.52 | 106.62 | -291.15 |
| LFU | Random | -178.68 | 84.33 | -261.76 | -155.80 | 101.88 | -257.68 |
| | MFU | **-180.53** | **72.54** | **-253.07** | -161.30 | 88.60 | -249.90 |
| LVF | Random | -177.85 | 78.25 | -256.10 | **-202.90** | **126.62** | **-219.68** |
| | MFU | -201.62 | 86.21 | -287.83 | -173.65 | 104.19 | -277.84 |
| Novel | Random | -206.41 | 127.08 | -333.62 | -186.05 | 113.12 | -299.18 |
| | MFU | -185.725 | 105.54 | -291.27 | -173.56 | 101.79 | -275.35 |

Using our parameterized *DDPGAgn*, the improvement of return over the redirector mode was 60.22% and 26.06% respectively using scalable and limited sized cache memory. But the retrieval costs have increased by 4.62% and 60.27% as well. The improvement in return is better than *StatAgn's* 47.94% using scalable memory [37]. We draw to the 'biasness to redirector mode' of the optimal policy as the reason. We identify the *DDPGAgn* to have reached some optimality concerning several features in these results: (i) all the metrics are better than that observed using the random policy (which with *ACAgn* was not the case), and (ii) the best configuration using scalable memory show a better earning net of penalties compared to no eviction policy, i.e., \$-109.16 > \$-116.95. We will indicate in Table 12 further the improvement in return using a sample of sequential results over the long run. As a result, we cannot accept our hypothesis that *DDPGAgn* will converge later than the *ACAgn.*

In order to investigate the reasons for the observed performance metrics, the key components of the returns need to be discussed. Figure 24 compares the total cost, penalty cost and retrieval costs as a proportion of the earnings among the best configurations in the short run. For instance, Penalty/Earning could be described as the penalty incurred to receive \$1 of earning from a context query response. It is a measure independent from the size of the query load. We use yellow to indicate Redirector mode, orange to indicate *StatAgn*, blue to indicate *ACAgn* and green to indicate *DDPGAgn* in the illustrations below.

Our cost optimization problem aims at minimizing the costs to earnings proportions which is evident with adaptive context caching in Figure 17. *StatAgn* is clearly advantageous in cost minimization among compared options but, it shows the highest *Penalty/Total Cost* ratio as well at 0.8389 and 0.8257 respectively using scalable cache (SC) and limited sized cache (LC). Yet, *DDPGAgn* using scalable cache memory is equally comparable. *ACAgn* using scalable cache, LVF eviction and MFU exploration shows the lowest of this ratio at 0.7131 followed by *DDPGAgn* using scalable cache, LFU eviction and random exploration at 0.7256. There are two possible reasons for this observation. Firstly, based on our assumption for PD and the divergence of number of evictions which we will indicate later in Figure 19, we can identify the penalty is reduced for RL agents as a result of gradually reaching an optimal policy. In contrast, *StatAgn* does not adjust or improve overtime. Secondly however, we could also attribute with the increase in retrieval costs of the RL agents caused by: (i) additional refreshing due to exploratory caching, and (ii) retrievals due to not caching relevant items as matter of not yet reaching the optimal policy.

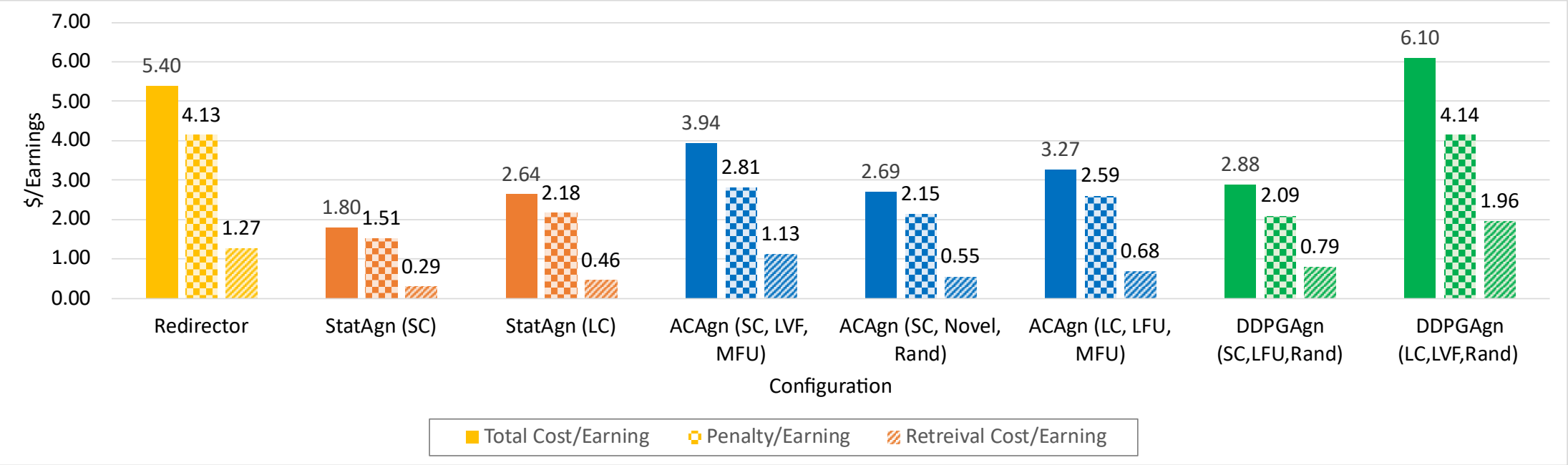

Figure 17: Comparison of *Per Earnings* for best configurations.

### *5.2.4Convergence and Long-term efficiency*

Further, we invested the performance of the selective caching agents over the long run. We executed each agent recurrently over thirty instances sequentially assuming the occurrence of the concerned *PlnPrd* for a month. But the timing of the queries was adjusted between each recurrence. This accumulates data and trains the model iteratively over time. We used the best configurations using scalable cache memory from the short run but set $\eta = 0.5$ based on the minimizing behaviour observed above. Figure 18 illustrated the variation of *PessiRet* and HR over the sequence of executions. Considering that the *PlnPrd* is relatively consistent across each consecutive day by definition, we investigated the impact of parameter sharing between executions for RL based agents. Provided we use a query load that is random in nature, we focused on sharing the $\varepsilon$ because could be used as an indicator for model convergence. We illustrate $\varepsilon$ sharing using triangle markers and round markers otherwise. *DDPGAgn* is excluded from Figure 18 due to redirector biasness of the learnt policy result in lower HR and *PessiRet* compared to the other agents. We indicate a sample of metrics over five consecutive recurrent executions of the *DDPGAgn* in order to illustrate this behaviour in Table 9.

Table 9: Average performance of agents over the long run

| Index | Earnings | Penalties | Retrieval Cost | HR | PessiRet |
|---|---|---|---|---|---|
| T+1 | 127.45 | 262.4 | 97.75 | 0.7288 | 0.0868 |
| T+2 | 126.45 | 255.5 | 92.15 | 0.7288 | 0.0868 |
| T+3 | 63.85 | 230.55 | 122.50 | 0.2295 | -1.2611 |

| Index | Earnings | Penalties | Retrieval Cost | HR | PessiRet |
|---|---|---|---|---|---|
| T+4 | 49.95 | 209.50 | 88.00 | 0.1694 | -1.127 |
| T+5 | 36 | 188.1 | 53.5 | 0.5375 | -0.3609 |

According to Figure 18, performance of the *StatAgn* in term of HR and *PessiRet* is marginally better than the RL based agents because of the random nature of the query load (lack of pattern) resulting in low learnability. We summarize the average of these results in Table 10. But there is a marginal decrease in *PessiRet* for *StatAgn* while it is relatively consistent for the *ACAgn*. Sharing $\varepsilon$ has caused a negative effect on the results but comparing against the nature of optimality in this case, it is acceptable. In fact, at the end of the repeated executions, $\varepsilon$=0.635 – a gradual increase in exploration. This is however unexpected as $\varepsilon$ should have reduced or maintained at a consistent value when the RL agents converge. We could identify two reasons in this case: (i) the set of sub-queries is not large enough to trigger many learning cycles; the average number of learning operations per recurrence is ~0.4 per minute, and (ii) lack of pattern in the query load resulting in the marginal average return of a window to be less than the $\text{Reward}_{min}$ threshold.

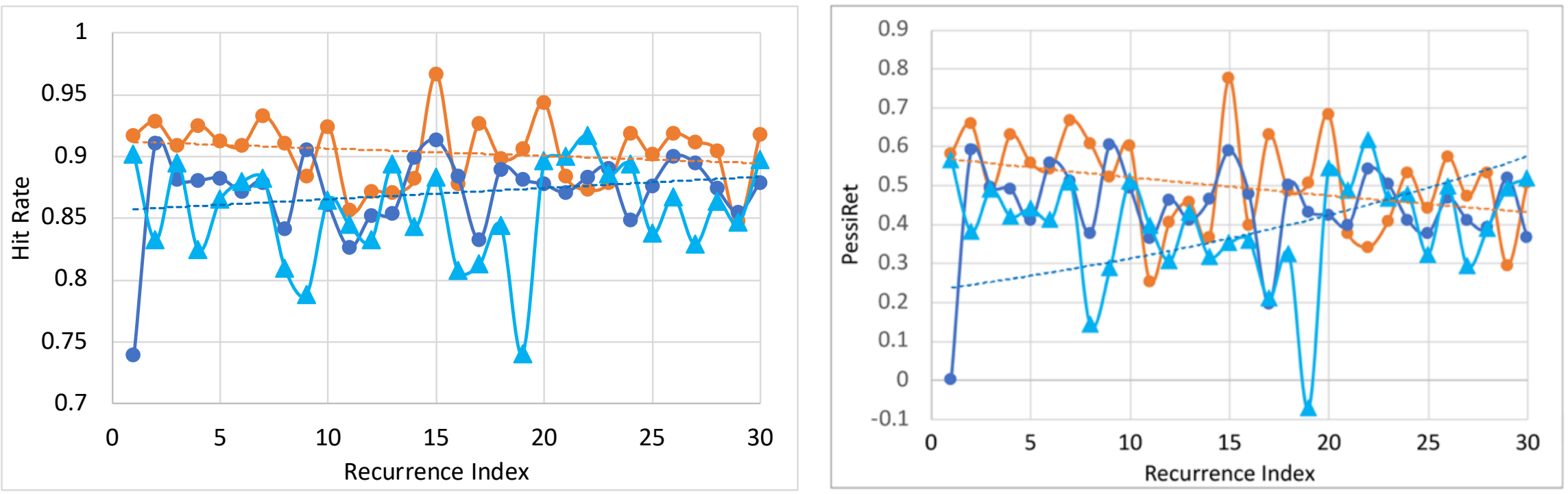

(a) (b)

Figure 18: Comparison (a) *PessiRet,* and (b) Hit Rate over long run.

Table 10: Average performance of agents over the long run

| Agent | $\varepsilon$-sharing | $\overline{HR}$ | $\overline{PessiRet}$ |
|---|---|---|---|
| StatAgn | False | $0.9036 \pm 0.0098$ | $0.5106 \pm 0.046$ |
| ACAgn | False | $0.8711 \pm 0.0123$ | $0.4416 \pm 0.044$ |
| | True | $0.8560 \pm 0.0150$ | $0.3975 \pm 0.052$ |
| DDPGAgn | False | $0.3170 \pm 0.2520$ | $-0.9210 \pm 0.623$ |
| | True | | |

Figure 19 illustrates the number of evictions in each repetition. There is an apparent reduction in evictions from cache for RL agents observed by the divergence in accumulated evictions. It indicates that "correct" entities and attributes has been cached which corresponds to our observation to converging to optimality overtime. Interestingly, sharing $\varepsilon$ for *ACAgn* and *DDPGAgn* result in performing worse than the *StatAgn* in entity evictions but marginally better in attribute evictions. It indicates that there exists no correlation between the entity and attribute evictions, proving our rationale to separate them under two logical levels (refer Figure 9) as a result of three reasons. Firstly, consider an entity that can be polymorphic, i.e., a vehicle can be a 'car' or a 'device' depending how the programmer define the entity in the query. The query referring to the vehicle as a device is only intermittent or less frequently accessed. Then the number of entity evictions would be independent from its related attributes if there are number of shared attributes such as location. Secondly, if an entity underpins multiple providers for the attributes, i.e., number of available slots originate from the car park system while the rating from a third-party API such as Google Maps, then it is possible only the entity reference be evicted given the provider's data is shared among other entities. Thirdly, it could be a result of entity duplication as well, i.e., $Park_2$ identified either by name or address, where attributes are cached in a shared manner in the ideal caching setting.

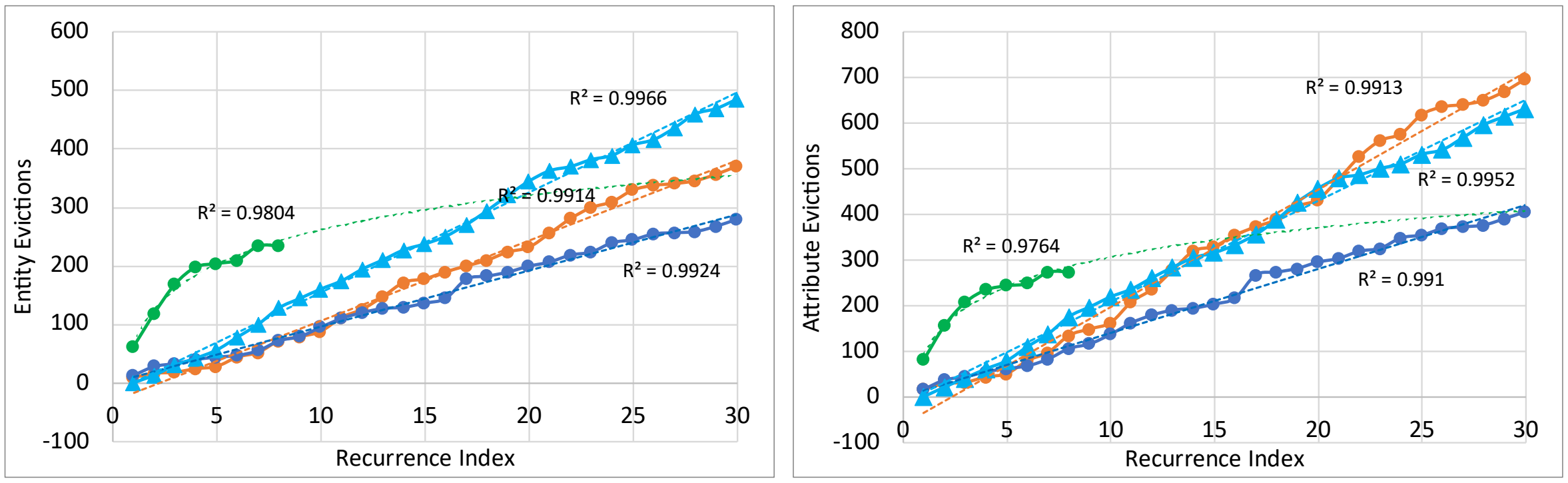

(a) (b)

Figure 19: Evictions of (a) entities, and (b) attributes over the long run.

### *5.2.5Key Observations*

In this section, we discuss the most important observations and identified principle for adaptive context caching. They can be listed as follows:

- Dynamically scaling cache memories are cost and performance effective. Consider the results using scalable cache for each selective caching agent. The overall return is greater and comparable against our benchmarks and consistency in QoS offering measured in variances are minimized using scalable context cache memories.
- The optimal adaptive context caching configuration is not universal in a distributed environment where cache memories are heterogenous, e.g., by size – limited sized, scalable.
- Solution for cost and performance effective adaptive context caching strategy is *pareto optimal* – the optimal policy $\pi^*$ could be one of several solutions. This could be observed by comparing the performance of *StatAgn* that show high HR and utilized cache memory space, i.e., scaling up to 9 entities in size (we can refer it being database mode bias), against the *DDPGAgn* which was redirector mode bias - lower hit rate and low utilized cache memory space, i.e., scaled only up to 6 entities in size.
- We theoretically argued in Section 3.3 that training the RL agent should be performed in a parameterized manner in order to adequately capture the volatility of the environment with the objectives of minimizing convergence time and maximize reliability. We tested our assumption comparing the two solutions: *ACAgn* and *DDPGAgn*. It is clear that *DDPGAgn* adapted much better than *ACAgn* in cost and performance efficiency, which confirms our assumption.
- The most important requirement of adaptive context caching (ACOCA) - ACOCA should be *long term optimal* but *short term aware*. The long-term schemes used in this work, i.e., *ACAgn* and *DDPGAgn* show better long-term cost optimality and resilience to noise however being poor in the short term, i.e., (i) transitioning from one PlnPrd to another, or (ii) when a significant change in the query load is observed. *StatAgn* performs on the contrary. It is important that the adaptive context caching incorporates both these aspects in order to maximize the profitability of the context management system. By "short term aware", we refer to four important considerations: (i) features of the query and context items, i.e., request rate of queries, refresh rate of attributes, (ii) co-occurrence, i.e., related queries that are requested in sequence such as $Q_5 \rightarrow Q_3$, inter-relations, i.e., weather attributes shared among $Q_4$ and $Q_5$, and intra-relationships among context items, i.e., entities involved in relational functions (*rFunction*) [13,14] such as *rFunction Ownership is Own (Person, Device)*, (iii) burstiness of the query load, i.e., sudden spike in request rate, that could be different from typically accessed context, e.g., alternate routes in the event of an accident, and (iv) characteristics of the query load, e.g., randomness. We intend to use this knowledge in our further work.

- In order to achieve the "*Logically-coherent*" objective of ACOCA-A, we identify that each component can be defined its primary objective derived from the above principle. For instance, the primary objective of cache memory module could be stated as: *Satisfying the QoS requirements for a stream of context queries* (short-term awareness) *whilst cache usage is optimized in a system-wide manner* (long-term optimality).

## 6 CONCLUSION AND FURTHER WORK

In our work, we introduced an approach for distributed adaptive context caching that optimizes for cost minimization and tested it as a proof-of-concept in this paper. We introduced two Reinforcement learning based approaches to make selective caching decisions that evaluates the profitability to cache an item "now". The paper presented several algorithms and heuristics to achieve our objective including eviction policies, and cache organization. These are explained and proved where necessary using mathematical models and our motivating scenario. We identify that the *StatAgn* and *DDPGAgn* have achieved our primary objective of cost optimization by up to 47.94% and 60.22% improvements in cost efficiency respectively against the redirector mode. We compared that with related work and established our approach is simplistic, computationally inexpensive, and much faster to converge compared to an RL based approach for the same overall improvement. We identified that LVF and TAH eviction policies performs well under highly volatile query loads where as LFU was suitable otherwise, especially with limited sized cache memory. The reason was due to the minimization of redundant costs such as hold-up costs and refreshes for intermittently accessed items. We showed that *StatAgn* is not suitable for solve the long-term cost optimization but rather reinforcement learning solutions are able to improve over time to cache only the most suitable items. RL based agents increase the returns and minimized evictions over the long-term. We establish using our results that caching context only on the basis of expecting low latency or cost as in data caching would not be effective. The optimal solution in the form of an optimal policy is pareto optimal comparing the result of *StatAgn* and *DDPGAgn*. Some context caching configurations perform only marginally better that the redirector mode which is the reason why context caching is much more non- trivial. Therefore, we finally derive a significant requirement of an adaptive context caching algorithm as: 'long term optimized whilst being short term aware'.

## ACKNOWLEDGMENTS

Support for this publication from the Australian Research Council (ARC) Discovery Project Grant DP200102299 is thankfully acknowledged.